\documentclass[letter,11pt]{article}
 
\usepackage{jheppub}

\usepackage{hyperref}
\usepackage{graphicx}
\usepackage{amsmath}
\usepackage{amssymb} 
\usepackage{xspace}
\usepackage{slashed}
\usepackage{subcaption}
\usepackage[normalem]{ulem}
\usepackage[dvipsnames]{xcolor}
\usepackage[utf8]{inputenc}
\usepackage[T1]{fontenc}
\usepackage{mathtools}
\usepackage{stmaryrd}
\usepackage{enumerate}
\usepackage{mathrsfs}

\newcommand{\nocontentsline}[3]{}
\newcommand{\tocless}[2]{\bgroup\let\addcontentsline=\nocontentsline#1{#2}\egroup}

\newcommand{\nn}{\nonumber}

\newcommand{\bea}{\begin{eqnarray}}
\newcommand{\eea}{\end{eqnarray}}

\def\ln{\textrm{ln}}

\def\nn{\nonumber}

\allowdisplaybreaks[2]

\usepackage{marginnote}

\preprint{\begin{flushright}
MIT--CTP 5268
\end{flushright}}

\title{Forward scattering in a thermal plasma}

\author{Varun Vaidya}
\affiliation{Center for Theoretical Physics, Massachusetts Institute of Technology, Cambridge, MA~02139, U.S.A.}

\emailAdd{vvaidya@mit.edu}

\abstract{I examine the regime of forward scattering of an energetic particle in a Plasma medium in thermal equilibrium. Treating the particle as an open quantum system interacting with a bath, I look at the time evolution of the reduced density matrix of the system. The kinematic and dynamical time scales that emerge can exist in several possible hierarchies which can lead to different EFT formulations. I show that in certain hierarchies, it becomes necessary to account for arbitrary number of coherent exchanges between the system and the bath going beyond the independent scattering paradigm. Analytic results are obtained in certain limits and the formalism is applied for the measurement of transverse momentum broadening of a quark in a Quark Gluon Plasma medium.
}

\begin{document}
\maketitle

\section{Introduction}

The Quark Gluon Plasma(QGP) is a state of quasi free quarks and gluons that existed in the early universe and has been recently created in heavy ion collision experiments at both RHIC and LHC. The individual nucleons in the heavy ions undergo energy stopping collisions which raises the temperature sufficiently for the quarks and gluons to be deconfined for a brief period of time forming the QGP medium. This droplet of QGP eventually expands and cools forcing the partons once again into color neutral hadrons which are captured by the detector. \\
Apart from energy stopping collisions that create the QGP, there are background hard events that create partons with energy far greater than the QGP temperature. These partons which evolve into jets may travel through a large portion of the QGP medium, thus getting significantly modified as compared to a pure vacuum evolution. The idea then is to exploit these jets as hard probes of the QGP medium and relate the modification of the properties of the jet such as jet energy, transverse momentum broadening, or more involved jet substructure observables to the properties of the medium. To carry out this plan therefore requires a reliable theoretical prediction for jet observables in a QGP background. 

A phenomenon that has been extensively studied in  literature\cite{Gyulassy:1993hr,Wang:1994fx,Baier:1994bd,Baier:1996kr,Baier:1996sk,Zakharov:1996fv,Zakharov:1997uu,Gyulassy:1999zd,Gyulassy:2000er,Wiedemann:2000za,Guo:2000nz,Wang:2001ifa,Arnold:2002ja,Arnold:2002zm,Salgado:2003gb,Armesto:2003jh,Majumder:2006wi,Majumder:2007zh,Neufeld:2008fi,Neufeld:2009ep} is that of Jet quenching, which entails a suppression of particles with high transverse momenta in the medium. This has also been recently observed in experiments at both Relativistic Heavy Ion Collider (RHIC) \cite{Arsene:2004fa,Back:2004je,Adams:2005dq,Adcox:2004mh} and Large Hadron Collider (LHC) \cite{Aad:2010bu,Aamodt:2010jd,Chatrchyan:2011sx}. The suppression mechanism happens through the mechanism of energy loss when jets travel through the hot medium. The key to understand jet quenching and jet substructure modifications in heavy ion collisions is to understand how the jet interacts with the expanding medium. 

The dominant regime of interaction of an energetic parton with the medium was found to be the forward scattering regime. The quark energy is the hard scale Q while the medium is made up of soft partons with energy of the order of the QGP temperature T<< Q.  Apart from the kinematic and measurement scales, the interaction of the system with the medium yields an emergent scale, namely the inverse interaction time of the system with the medium $t_I$. Depending on $\alpha_s$ this scale can be comparable or hierarchically separated from the correlation time scale of the thermal bath $ t_e \sim 1/T$. We also have the time scale t which is the time of propagation of the jet in the medium which can also be treated as the length of the medium. The interaction of the partons in the thermal bath yields a medium induced gluon mass $ m_D\sim gT$. Finally we have the emergent confinement scale $\Lambda_{QCD}$ where we can no longer apply perturbation theory. \\
We always assume that the hard scale Q is much larger than all the other scales in the system which together form our Infra-Red or IR scales. However, a large hierarchy of scales can exist amongst the IR scales and would therefore lead to different expansion parameters in our EFT formulation. In this paper, I examine all possible hierarchical scenarios between the IR scales and derive expressions for the observable of interest, the transverse momentum broadening of a quark in the QGP medium. 

Due to the multiple scales involved in the problem, a powerful tool to deal with large hierarchies of scales is Effective Field Theory (EFT). The EFT that is extensively used for jet studies at high energy colliders is Soft-Collinear Effective Theory (SCET). There are also formulations of SCET (known as SCET$_G$) treating the Glauber gluon, which is a type of mode appearing in forward scattering, as a background field induced by the medium interacting with an energetic jet. By making use of the collinear sector of the corresponding  EFT, this formalism has been used to address the question of jet quenching in the medium \cite{Ovanesyan:2011kn,Chien:2015hda,Ovanesyan:2011xy,Chien:2015vja,Kang:2014xsa}. I will use a complimentary approach using a new EFT for forward scattering that has been developed recently \cite{Rothstein:2016bsq} which also uses the Glauber mode to write down contact operators between the soft and collinear momentum degrees of freedom which is ideally suited for the situation we want to study.

This EFT formulation was adopted for computing the transverse momentum broadening of the quark $\vec{q}_T \sim T$  in \cite{Vaidya:2020lih} using the scattering angle ($\theta \sim T/Q$ )as the expansion parameter. Here T is the QGP temperature. This was subsequently expanded to include the hard interaction creating the energetic parton in \cite{Vaidya:2020lih}, where a factorization formula for jet substructure observables was derived. In both these cases, the Lindblad equation was used to resum multiple interactions of the jet with the medium, which only accounts for independent scattering diagrams. In this paper, I would like to explore all the classes of Glauber exchanges going beyond the independent scattering paradigm.
Depending on the hierarchy of time scales, different classes of diagrams become relevant at each order in perturbation theory. In this paper, I only consider the corrections from Glauber mode exchanges which mediate forward interaction between a soft medium and a collinear jet. The radiative corrections from soft and collinear modes will be considered elsewhere. I compute the contribution of the Glauber exchange diagrams to the reduced density matrix of my jet and focus on the transverse momentum broadening observable which was also considered in \cite{Vaidya:2020cyi}. 
In a future work, this will eventually be incorporated into the factorization formula for jet substructure observables. 

This paper is organized as follows. Section \ref{sec:Forward} introduces the EFT in the forward scattering regime and its application to the evolution of the reduced density matrix of an energetic parton in the thermal medium. Section \ref{sec:GExc} looks at Glauber exchange diagrams order by order in perturbation theory and their scaling in terms of various time scales. This is followed by an all order expression for the system interaction time ($t_I$) in Section \ref{sec:AllOrder}. Section \ref{sec:RDen} examines the contribution of the kernel $t_I$ to the reduced density matrix followed by possible simplifications in various limits in Section \ref{sec:Simp}. Section \ref{sec:Con} gives a summary and an outline for future improvements.  


\section{EFT in the forward scattering regime}
\label{sec:Forward}

As motivated in \cite{Vaidya:2020cyi}, the dominant regime of interaction between energetic partons and the thermal medium is forward scattering where the cross section for 2 $\rightarrow $ 2 scattering is inversely proportional to the $4^{th}$ power of the scattering angle $\theta$. We therfore utilize the EFT for forward scattering developed in \cite{Rothstein:2016bsq} with $\lambda =\theta <<1$ as the expansion parameter. This EFT is developed within the framework of Soft collinear Effective Theory~\cite{Bauer:2001ct,Bauer:2000yr,Bauer:2002nz} where the t channel forward scattering is mediated by Glauber modes. The jet is made up of collinear partons while the medium is composed of soft modes. In light cone-co-ordinates, these momenta scale as 
\begin{align}
 p_c^{\mu} \sim Q( 1, \lambda^2, \lambda ) \ \ \ \ \text{collinear } \nn\\
 p_s^{\mu} \sim Q( \lambda, \lambda, \lambda) \ \ \ \text{Soft} \nn\\
p_G^{\mu} \sim Q( \lambda, \lambda^2, \lambda) \ \ \ \text{Glauber} 
\end{align}

where we have decomposed momenta as $p^{\mu} \equiv ( \bar n\cdot p, n \cdot p,  \vec{p}_{\perp})$ with $ n^{\mu} \equiv (1,0,0, 1) $ and $\bar n^{\mu} \equiv (1,0,0,-1)$. This assumes without loss of generality that the collinear partons are moving along the z direction.

The Glauber mode is non-propagating and scales in such a way that a Glauber absorption/emission maintains the momentum scaling of the Collinear and Soft modes. 

We want to describe the transverse momentum broadening of a quark as it traverses a region of a QGP medium in thermal equilibrium over time t. 
This is done by treating the jet as an open quantum system interacting with a thermal bath. We can trace over the Soft thermal degrees of freedom and follow the the time evolution of the reduced density matrix of the jet. As in \cite{Vaidya:2020cyi}, we look at a diagonal element of the density matrix which gives us the transverse momentum distribution of the quark.  We start with an initial state density matrix 
\begin{align}
\rho(0) = |Q\rangle \langle Q| \otimes \rho_B 
\end{align}
where $|Q \rangle$ is a collinear quark states and $\rho_B$ is the thermal density matrix of the QGP medium which we assume to be independent of time and is initially unentangled with the collinear quark. We can write the effective Hamiltonian as 
\begin{align}
H = H_c+H_s+ H_{G}  
\end{align}

The $H_c$ and $H_s$ Hamiltonians have a free particle evolution and also induce interactions among the collinear and Soft partons respectively. 
The effective gauge invariant operators which make up the Glauber interaction Hamiltonian $H_G$ for quark-quark ($qq$), quark-gluon ($qg$ or $gq$) and gluon-gluon ($gg$) interactions have been worked out in the Feynman gauge in Ref.~\cite{Rothstein:2016bsq}
\begin{align}
\label{EFTOp}
\mathcal{O}_{ns}^{qq}&=\mathcal{O}_n^{qB}\frac{1}{\mathcal{P}_{\perp}^2}\mathcal{O}_s^{q_nB} \,, \ \ \ \ \ \mathcal{O}_{ns}^{qg}=\mathcal{O}_n^{qB}\frac{1}{\mathcal{P}_{\perp}^2}\mathcal{O}_s^{g_nB} \,, \nn\\
\mathcal{O}_{ns}^{gq}&=\mathcal{O}_n^{gB}\frac{1}{\mathcal{P}_{\perp}^2}\mathcal{O}_s^{q_nB} \,, \ \ \ \ \ \mathcal{O}_{ns}^{gg}=\mathcal{O}_n^{gB}\frac{1}{\mathcal{P}_{\perp}^2}\mathcal{O}_s^{g_nB}\,,
\end{align}
where $B$ is the color index and the subscripts $n$ and $s$ denote the collinear and soft operators respectively. The soft operators $\mathcal{O}_s$ are constructed from the gauge invariant soft quark and gluon building blocks that are built out of the soft fields dressed with soft Wilson lines:
\begin{align}
\mathcal{O}_s^{q_nB} &= 8\pi \alpha_s\left(\bar \psi^n_s T^B \frac{\slashed{n}}{2}\psi_s^n\right) \,,\nn\\
\psi_s^n &= S_n^{\dagger}\psi_s \,, \nn\\ 
\mathcal{O}^{g_nB}_s &= 8\pi \alpha_s\left(\frac{i}{2}f^{BCD}\mathcal{B}_{s\perp }^{nC}\frac{n}{2}\cdot (\mathcal{P}+\mathcal{P}^{\dagger}) \mathcal{B}_{s\perp}^{nD} \right) \,, \nn\\
\label{eqn:softO}
\mathcal{B}_{s\perp}^{n\mu} &= \mathcal{B}_{s\perp}^{nB\mu} T^B =  \frac{1}{g}\big(S_n^{\dagger}iD_{s\perp}^{\mu} S_n\big) \,,
\end{align}
where the soft Wilson lines ensure that the operators are invariant under soft gauge transformations.

The collinear operators are built out of the collinear building blocks. In this paper, we will only work with collinear quarks which are constructed from bare collinear quark fields dressed with collinear Wilson lines:
\begin{align}
\mathcal{O}_n^{qB} &= \bar{\chi}_n T^B \frac{\slashed{\bar n}}{2} \chi_n \\
\chi_n &= W_n^\dagger \xi_n = W_n^\dagger \frac{\slashed{n}\slashed{\bar{n}}}{4}\psi \,,
\end{align}
where $\psi$ is the standard four-component Dirac spinor.
Since the soft momentum puts the collinear particle off-shell and off-shell modes have been integrated our in the construction of the EFT, the collinear fields do not transform under the soft gauge transformations.

In this paper we will ignore any self interactions in the Soft and collinear sectors and concentrate on the all orders evolution induced by just the Glauber Hamiltonian. Therefore, the time evolution for my density matrix is simply 
\begin{align}
\rho(t) = \mathcal{T}\Big\{e^{-i\int_0^t dt' H_{G,I}(t')}\Big\}\rho(0)\bar {\mathcal{T}}\Big\{ e^{-i\int_0^t dt' H_{G,I}(t')}\Big\}
\label{DEvol}
\end{align}
where $H_{G,I}$ is the Glauber Hamiltonian in the interaction picture. The matrix element we want to look at is $\langle q| \rho(t) |q\rangle $
which can be evaluated by expanding out the Glauber Hamiltonian to each order as a perturbation.
Depending on the hierarchy of time scales, we can then look at the class of diagrams that becomes important in each scenario and whether an all order resummation is required. Since the trace over the density matrix is to be preserved, we can choose to look only at those diagrams which have atleast one Glauber insertion on each side of the cut($\Sigma_R(t)$). The contribution from the diagrams which have no Glauber insertions on one side of the cut($\Sigma_V(t)$) can then be inferred from preservation of probability. 
\begin{align}
\langle q| \rho(t) |q\rangle = \langle q| \rho(0) |q\rangle + \Sigma_V(t)+ \Sigma_R(t) 
\label{Den}
\end{align}

We now consider contribution from $\Sigma_R(t)$ that appear at each order in perturbation and compute the corrections for different approximations.


\section{Glauber Ladder exchanges in thermal field theory}
\label{sec:GExc}

To compute the diagrams, we will work in the real time formalism of thermal field theory. The Soft degrres of freedom make up the medium at thermal equilibrium at temperature T. To use the real time formalism involves doubling of the field content for the soft degrees of freedom. Since we will not be computing any corrections from insertion of the Soft Hamiltonian in this paper, hence we need not be concerned about the doubling of the fields and we can work with a single copy. The only difference would be that we need to replace the vacuum time ordered as well as on-shell propagators with the corresponding thermal propagators.\\

So we have 
\begin{align}
\frac{i}{k^2+i\epsilon} \rightarrow \frac{i}{k^2+i\epsilon}- 2\pi \delta(k^2)n(|k^0|)\nn\\
2\pi \delta^+(k^2) \rightarrow  2\pi \delta^+(k^2) - 2\pi \delta(k^2)n(|k^0|)
\end{align}
where 
\begin{align}
n(|p^0|) = \frac{\theta(p^0)}{e^{\beta p^0}+1} + \frac{\theta(-p^0)}{e^{-\beta p^0}+1} 
\end{align}
which is valid for fermions. I will only consider Collinear Quark-Soft quark interactions in this paper, which are easily generalized for the other cases.

We now look at the real diagram that contribute to $\Sigma_R(t)$ by expanding out the Glauber Hamiltonian in the time evolution of the density matrix as in Eq. \ref{DEvol}. 


\subsection{ $O(H_G^2)$} 

\label{sec:TwoG}

At this order, we have a single diagram with one Glauber exchange on each side of the cut as shown in Fig.\ref{Hg2}. The black dashed line is the collinear fermion  while the green one is the Soft thermal fermion. The Glauber is indicated by the red dotted line.
	
	\begin{figure}[h!]
   \begin{minipage}{.25\textwidth}
    \centering
    \includegraphics[width=0.8\textwidth]{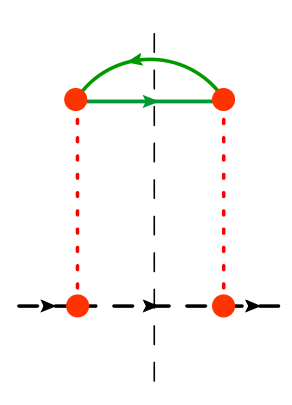}
    \caption{}
		\label{Hg2}
  \end{minipage}
	 \begin{minipage}{.7\textwidth}
    \begin{align}
      \Sigma_R^{(1)} &= |C_{qq}|^2\int d^4x_1 \int d^4x_2  \langle X_s| \frac{1}{\mathcal{P}_{\perp}^2}O^A_s(x_1) \rho_B  \frac{1}{\mathcal{P}_{\perp}^2}O^B_s(x_2)|X_s\rangle \nn\\
			&\langle q| O_n^A(x_1)\rho_n(0) O_n^B(x_2)|q\rangle 
			\label{HG2}
    \end{align}
		 \end{minipage}%
\end{figure}
	
	The Glauber operator used here is $H_G \equiv O_{ns}^{qq}= C_{qq}O_n^A 1/\mathcal{P}_{\perp}^2 O_s^A$ which mediates the interaction between a soft and a collinear quark. $C_{qq}$ is simply the Wilson coefficient for this operator which at tree level is $-2ig^2$. However, I will not be too careful about overall factors in this section since our objective is to deduce the scaling behavior of the corrections. This will be taken care of in a later section \ref{sec:AllOrder} where we present the all orders result. The contribution to $\Sigma_R(t)$ in Eq.\ref{Den} at this order is given by Eq. \ref{HG2}.
	
	The collinear sector matrix element gives us 
	\begin{align}
	\langle q| O_n^A(x_1)\rho_n(0) O_n^B(x_2) |q\rangle  &= \int \tilde dp_1\tilde dp_2 e^{i x_1 \cdot(q-p_1)}e^{-ix_2 \cdot(q-p_2)}\text{Tr}\Big[\bar u(q)T^A \frac{\slashed{\bar n}}{2}u(p_1) \bar u(p_2) T^B \frac{\slashed{\bar n}}{2} u(q)\Big]\langle p_1| \rho(0) |p_2\rangle \nn
	\end{align}
	
	where the phase space integral measure is 
	\begin{align}
	\tilde dp= \frac{d^3p}{(2\pi)^3 2E_p} = \frac{d^4p}{(2\pi)^3} \delta^+(p^2)
	\end{align}
The Soft sector matrix element, which is obtained by doing a trace over all Soft states $|X_s\rangle$ gives us 
\begin{align}
 &\langle X_s| \frac{1}{\mathcal{P}_{\perp}^2}O^A_s(x_1) \rho_B  \frac{1}{\mathcal{P}_{\perp}^2}O^B_s(x_2)|X_s\rangle = -2\int \frac{d^4 k}{(2\pi)^4} \frac{1}{k_{\perp}^2}\frac{1}{k_{\perp}^2} e^{-ik \cdot (x_1-x_2)}\text{Tr}\Big[T^AT^B\Big]\nn\\
&\times \int \frac{d^4p}{(2\pi)^2} \left(\delta^+(p^2)- \delta(p^2)n(|p^0|)\right) \left((\delta^+(-p+k)^2-  \delta((-p+k)^2)n(|-p^0+k^0|)\right)(p^+)^2\nn\\
&\equiv  \int \frac{d^2k_{\perp}dk^+}{(2\pi)^4} \frac{1}{k_{\perp}^2}\frac{1}{k_{\perp}^2} e^{-ik \cdot (x_1-x_2)} D^{AB}_>(k)
\end{align}
where we can define $D^{AB}_>(k) $, the advanced Wightman function, with $p^+ =n \cdot p$. 
  Given the color structure for this function, we can write 
\begin{align}
 D^{AB}_>(k) = \delta^{AB} T_F D_>(k)
\label{Wight}
\end{align}
	Finally, we can simplify the Wightman function as 
	
	\begin{align}
	D_>(k) &= -\int dk^-\int \frac{d^4p}{(2\pi)^2} \left(\delta^+(p^2)- \delta(p^2)n(|p^0|)\right) \left((\delta^+(-p+k)^2- \delta((-p+k)^2)n(|-p^0+k^0|)\right)(p^+)^2\nn\\
	&= \int \frac{d^2p_{\perp}}{2(2\pi)^2} dp^+ \Big\{ n_F\left(|\frac{p^+}{2}+\frac{(\vec{p}_{\perp}+\vec{k}_{\perp})^2}{2p^+}|\right) \left(1- n_F\left(|\frac{p^+}{2}+\frac{(\vec{p}_{\perp})^2}{2p^+}|\right)\right) \nn\\
	&+ n_F\left(|\frac{p^+}{2}+\frac{p_{\perp}^2}{2p^+}|\right) \left(1- n_F\left(|\frac{p^+}{2}+\frac{(\vec{p}_{\perp}+\vec{k}_{\perp})^2}{2p^+}|\right)\right)\Big\}\Theta(p^+)
	\end{align}
	
Combine the collinear and soft sectors, and doing the integral over spatial co-ordinates yields 3-momentum conserving $\delta$ functions $\delta^3(q-p_1-k)\delta^3(q-p_2-k)$

 that  set $\vec{p}_1= \vec{p}_2$ and so $p_1^0 =p_2^0 \equiv p^0 =|\vec{q}-\vec{k}|$. We can eliminate the integrals over $p_1, p_2$ 

\begin{align}
 \Sigma_R^{(1)}&=|C_{qq}|^2T_F\int \frac{d^2k_{\perp}dk^+}{(2\pi)^4} \frac{1}{k_{\perp}^4}e^{-ik^0(t_1-t_2)} D_>(k)\int_0^t dt_1 \int_0^t dt_2\frac{1}{4p_0^2} e^{i t_1 (q^0-p^0)}e^{-it_2 (q^0-p^0)}\nn\\
	&\times\text{Tr}\Big[\bar u(q)T^A \frac{\slashed{\bar n}}{2}u(p) \bar u(p) T^A \frac{\slashed{\bar n}}{2} u(q)\Big]\langle p| \rho(0) |p\rangle \nn
	 \end{align}

Now we can do the integral over $t_1, t_2$ in the limit $t \rightarrow \infty$.  This assumes that the time scales of propagation of the quark in the medium is much larger than the inverse of the energy of all the degrees of freedom in play. Then using 
\begin{align}
\text{limit}_{t\rightarrow \infty} \int_0^t dt_1 \int_0^t dt_2 e^{i \omega t_2}e^{-i \omega t_1} = 2\pi t\delta(E)
\end{align}
we now have 
\begin{align}
 \Sigma^{(1)}&=t |C_{qq}|^2T_F\text{Tr}\Big[\bar u(q)T^A \frac{\slashed{\bar n}}{2}u(p) \bar u(p) T^B \frac{\slashed{\bar n}}{2} u(q)\Big]\langle p| \rho(0) |p\rangle \nn\\
	 &\times\int \frac{d^2k_{\perp}dk^+}{(2\pi)^3} \frac{1}{k_{\perp}^2}\frac{1}{k_{\perp}^2} D^{AB}_>(k)\delta(|\vec{q}-\vec{k}|-q^0-k^0)\frac{1}{4(q^0-k^0)^2}
\end{align}
We can write the delta function in a covariant manner and then apply power counting, 
\begin{align}
\delta(|\vec{q}-\vec{k}|-q^0-k^0) \approx  q^-\delta^+((q^+-k^+)-\frac{(\vec{q}_{\perp}-\vec{k}_{\perp})^2}{q^-})
\end{align}

	We can finally write, performing the average over initial state spin and color.
	\begin{align}
	 \Sigma^{(1)} &= |C_{qq}|^2T_FC_F\times t\times \int \frac{d^2k_{\perp}}{(2\pi)^3}\langle q-k|\rho(0) |q-k\rangle \frac{1}{(k_{\perp}^2)^2} D_>(k) \nn\\
	& \equiv  t \int d^2 k_{\perp} F(k_{\perp} )  \langle q-k| \rho(0) |q-k\rangle
	\label{Hg2R}
	\end{align}
	which agrees with the result obtained in \cite{Vaidya:2020cyi}  using the Imaginary time formalism.
	This expression leads to the emergence of a dynamical time scale namely the typical time scale for the interaction of the system with the medium. 
	\bea
	t_I = \left( \int d^2 k_{\perp} F(k_{\perp} ) \right)^{-1}
	\eea
	
	In general, this time scale will receive higher order corrections in $\alpha_s$. If we assume that the measurement scale $q_{\perp}$ that we impose in our quark is perturbative, then higher order corrections to this time scale will be suppressed by perturbative coupling and hence the term that we have derived  will be the dominant term. The closer our transverse momentum gets to $\Lambda_{QCD}$, the closer $\alpha_s$ gets to 1 and then it becomes necessary to include higher order corrections to the time scale, a scenario which we will consider in later sections. \\
Lets first focus on the case when $t_I$ is perturbatively calculable and the first term suffices. 
In this case we can examine three possibilities 
\begin{itemize}
\item{$\lambda_t <<1 $}
In this case the term we have computed suffices as a good approximation. We can then write the final result for our reduced density matrix element as 
\bea
\langle q| \rho(t) |q\rangle = \langle q| \rho(0) |q\rangle- t \langle q| \rho(0) |q\rangle\int d^2 k_{\perp} F(k_{\perp} )  +  t \int d^2 k_{\perp} F(k_{\perp} )  \langle q-k| \rho(0) |q-k\rangle \nn
\eea

\item{$\lambda_t \sim 1$} 

we can have a scenario such that $\lambda_t= t/t_I \sim 1 $, where t is the time of propagation in the medium. In this case it becomes necessary to resum all corrections with powers of $\lambda_t$, if any such terms are present.
\item{$\lambda_t >>1$}

In this case, since t is so large, the perturbative corrections to $ t_I$, even if they are suppressed by $\alpha_s$, give an O(1) contribution. The perturbative order to which we need to compute $t_I$ would then depend on the magnitude of t. 

\end{itemize}

We can now look at the higher order correction coming from Glauber Gluon exchanges. The objective would be identify the type of corrections that we see and decide which correction is important in what hierarchy of scales.  At the same time, we can try and infer the structure for this type( Glauber gluon) of correction to all orders in perturbation theory.


\subsection{O($H_G^4$)}
	
	Since  a single Glauber exchange leads to a phase factor, we expect that diagrams at the level of amplitude squared with odd number of Glauber exchanges will vanish when combined with their complex conjugate result. 
	In order to gain intuition about higher order corrections, which can thereafter be generalized, I therefore look at the O($H_G^4$) Glauber exchange. 
 We have two possibilities here, namely two-two Glauber insertions  vs three-one Glauber insertions across the cut. Again, we will consider only the 2-2 case since the objective here is to establish the scaling for different diagrams and the 2-2 case serves as an adequate example. We will deal with the other possibilities  in detail in section \ref{sec:AllOrder} when we examine the all order structure for Glauber exchanges. For the 2-2 Glauber insertion, we can have a box as well as a cross box topology and we will consider each one on turn. The contribution to the reduced density matrix becomes  

	\bea
	\Sigma_R^{(2)} &=& \frac{(-i)^4}{(2!)^2} \langle q X_s|\int d^4x_1 d^4x_2 T\Big\{H_{G,I}(x_1) H_{G,I}(x_2)\Big\}\rho(0) \int d^4y_1 d^4y_2\bar T\Big\{H_{G,I}(y_1) H_{G,I}(y_2)\Big\} |q X_s\rangle \nn\\
	&=&  |C_{qq}|^4\frac{(-i)^4}{(2!)^2} \int d^4x_1 d^4x_2 \int d^4y_1 d^4y_2\langle q| T\Big\{O_n^{A_1}(x_1) O_n^{A_2}(x_2)\Big\}\rho_n(0) \bar T\Big\{O_n^{B_1}(y_1) O_n^{B_2}(y_2)\Big\} |q\rangle \nn\\
	&\times& \langle X_s| T\Big\{\frac{1}{\mathcal{P}_{\perp}^2}O_s^{A_1}(x_1)\frac{1}{\mathcal{P}_{\perp}^2} O_s^{A_2}(x_2)\Big\}\rho_B \bar T \Big\{\frac{1}{\mathcal{P}_{\perp}^2}O_s^{B_1}(y_1)\frac{1}{\mathcal{P}_{\perp}^2} O_s^{B_2}(y_2)\Big\} |X_s\rangle
		\eea

	We first look at the collinear sector matrix element 
	\bea
	&&\langle q| T\Big\{O_n^{A_1}(x_1) O_n^{A_2}(x_2)\Big\}\rho_n(0) \bar T\Big\{O_n^{B_1}(y_1) O_n^{B_2}(y_2)\Big\} |q\rangle\nn\\
	 &=& 4\int \tilde dp_1 \tilde dp_2 \bar u(q)e^{iq \cdot x_1}\frac{\slashed{\bar n}}{2}T^{A_1}\int \frac{d^4l_1}{(2\pi)^4}\frac{e^{-il_1\cdot (x_1-x_2)}i\slashed{l}}{l_1^2+i\epsilon}\frac{\slashed{\bar n}}{2}T^{A_2}u(p_1)e^{-ip_1\cdot x_2}\langle p_1|\rho_n(0)|p_2\rangle \nn\\
	&\times&\bar u(p_2)e^{ip_2 \cdot y_2}\frac{\slashed{\bar n}}{2}T^{B_2}\int \frac{d^4l_2}{(2\pi)^4}\frac{e^{il_2\cdot (y_1-y_2)}(-i)\slashed{l}}{l_2^2-i\epsilon}\frac{\slashed{\bar n}}{2}T^{B_1}u(q)e^{-iq\cdot y_1} 
	\label{CollHG4}
	\eea

	In this case, we cannot simplify any further and must first look at the Soft sector traced over all soft final states 
	\bea
S^{(4)}	&=& \langle X_s| T\Big\{\frac{1}{\mathcal{P}_{\perp}^2}O_s^{A_1}(x_1)\frac{1}{\mathcal{P}_{\perp}^2} O_s^{A_2}(x_2)\Big\}\rho_B \bar T \Big\{\frac{1}{\mathcal{P}_{\perp}^2}O_s^{B_1}(y_1)\frac{1}{\mathcal{P}_{\perp}^2} O_s^{B_2}(y_2)\Big\} |X_s\rangle 
		\eea
			There are several possible diagrams that can be drawn here, and we need to see which of them gives the dominant contribution for a given hierarchy of time scales. One of then corresponds to independent scattering centers while the other involves correlated scattering centers in the medium. The independent case has two topologies, planar and non-planar and we will consider each one in turn

		\begin{itemize}
		\item{ Independent scattering centers} 		

			\begin{figure}[h!]
   \begin{minipage}{.25\textwidth}
    \centering
    \includegraphics[width=0.9\textwidth]{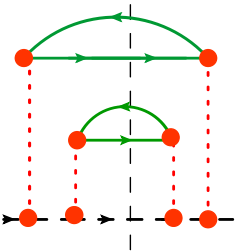}
    \caption{Planar}
		\label{Hg2}
  \end{minipage}
	 \begin{minipage}{.7\textwidth}
    \begin{align}
     &S^{(4)}_{I}=\langle X_s| \frac{1}{\mathcal{P}_{\perp}^2}O_s^{A_1}(x_1)\frac{1}{\mathcal{P}_{\perp}^2} O_s^{B_1}(y_1)\rho_B  |X_s\rangle \nn\\
		&\langle X_s| \frac{1}{\mathcal{P}_{\perp}^2}O_s^{A_2}(x_2)\frac{1}{\mathcal{P}_{\perp}^2} O_s^{B_2}(y_2)\rho_B |X_s\rangle \nn\\
		\label{HGIn}
		 \end{align}
		 \end{minipage}%
\end{figure}

			 The soft matrix element for planar independent scattering is given in Eq.\ref{HGIn}. This effectively gives us the product of two correlators in the thermal medium, each of which are identical to the single scattering case considered above. 
	\begin{align}
		S_I^{(4)}	&= \int \frac{d^2k_{1,\perp}dk_1^+}{(2\pi)^4}\frac{e^{-ik_1\cdot(x_1-y_1)}}{(k_{1,\perp})^4}D_>^{A_1B_1} \int \frac{d^2k_{2,\perp}dk_2^+}{(2\pi)^4}\frac{e^{-ik_2\cdot(x_2-y_2)}}{(k_{2,\perp})^4}D_>^{A_2B_2}
		\end{align}
		where $D_>^{AB}$ is defined in Eq. \ref{Wight}.	Now combining this with the collinear sector  matrix element Eq.\ref{CollHG4} and performing the spatial integrals yields			
		\bea
		(2\pi)^3 \delta^3(q-l_1-k_1)(2\pi)^3 \delta^3(-p_1+l_1-k_2)(2\pi)^3 \delta^3(l_2-q+k_1)(2\pi)^3 \delta^3(p_2-l_2+k_2)\nn
		\eea
	The momentum conservation implies $\vec{l}_1=\vec{l}_2$ and therefore $\vec{p}_1=\vec{p}_2 =\vec{q}-\vec{k}_1-\vec{k}_2\equiv \vec{p}$, we can then eliminate integrals over $\vec{l}_1,\vec{l}_2$ and $\vec{p}_1,\vec{p}_2$,
	
	\bea
		 \Sigma_R^{(2)}&=& |C_{qq}|^4\int_0^t dx_1^0 \int_0^t dx_2^0\int_0^t dy_1^0\int_0^t dy_2^0\nn\\
		&\times&\bar u(q)e^{iq^0 t_1}\frac{\slashed{\bar n}}{2}T^{A_1}\int \frac{dl_1^0}{2\pi}\frac{e^{-il_1^0(x_1^0-x_2^0)}i\slashed{l}}{(l_1^0)^2-(\vec{q}-\vec{k}_1)^2+i\epsilon}\frac{\slashed{\bar n}}{2}T^{A_2}u(p)e^{-ip_1^0 x^0_2}\langle p|\rho_n(0)|p\rangle \nn\\
	&\times&\bar u(p)e^{ip^0 y^0_2}\frac{\slashed{\bar n}}{2}T^{B_2}\int \frac{dl_2^0}{2\pi}\frac{e^{il_2^0 (y_1^0-y_2^0)}(-i)\slashed{l}}{(l_2^0)^2-(\vec{q}-\vec{k}_1)^2-i\epsilon}\frac{\slashed{\bar n}}{2}T^{B_1}u(q)e^{-iq^0 y_1^0} \nn\\
	&\times& \int \frac{d^2k_{1,\perp}dk_1^+}{(2\pi)^4}\frac{e^{-ik_1^0 (x_1^0-y_1^0)}}{(k_{1,\perp})^4}D_>^{A_1B_1} \int \frac{d^2k_{2,\perp}dk_2^+}{(2\pi)^4}\frac{e^{-ik_2^0 (x_2^0-y_2^0)}}{(k_{2,\perp})^4}D_>^{A_2B_2}\frac{1}{4(p^0)^2}\nn
			\eea
		
		The idea is that we should take the limit $t \rightarrow \infty$ at the very end, 
		To proceed then we must first do the contour integrals over $l_1^0, l_2^0$, There are two poles, but we are only interested in the one which leads to a t channel exchange and hence leads to the proper scaling for the Glaubers. The other pole leads to an s channel exchange that will result in a power suppressed result. So we only retain $l_1^0, l_2^0\geq 0$. For $l_1^0$, we therefore only consider the case $x_1^0 \geq x_2^0$, closing the contour in the lower half plane, while for $l_2^0$, we close the contour in the upper half of the plane, which retains the contribution $y_1^0 \geq y_2^0$, 
		\bea
		 \Sigma_R^{(2)}&=& |C_{qq}|^2\int_0^t dx_1^0 \int_0^t dx_2^0\int_0^t dy_1^0\int_0^t dy_2^0\Theta(x_1^0-x_2^0)\Theta(y_1^0-y_2^0)\nn\\
		&\times&\bar u(q)e^{iq^0 t_1}\frac{\slashed{\bar n}}{2}T^{A_1}\frac{e^{-i(|\vec{q}-\vec{k}_1|)(x_1^0-x_2^0)}i\slashed{l_1}}{2|\vec{q}-\vec{k}_1|}\frac{\slashed{\bar n}}{2}T^{A_2}u(p)e^{-ip^0 x^0_2}\langle p|\rho_n(0)|p\rangle \nn\\
	&\times&\bar u(p)e^{ip^0 y^0_2}\frac{\slashed{\bar n}}{2}T^{B_2}\frac{e^{i(|\vec{q}-\vec{k}_1|) (y_1^0-y_2^0)}(-i)\slashed{l_2}}{2|\vec{q}-\vec{k}_1|}\frac{\slashed{\bar n}}{2}T^{B_1}u(q)e^{-iq^0 y_1^0} \nn\\
	&\times& \int \frac{d^2k_{1,\perp}dk_1^+}{(2\pi)^4}\frac{e^{-ik_1^0 (x_1^0-y_1^0)}}{(k_{1,\perp})^4}D_>^{A_1B_1} \int \frac{d^2k_{2,\perp}dk_2^+}{(2\pi)^4}\frac{e^{-ik_2^0 (x_2^0-y_2^0)}}{(k_{2,\perp})^4}D_>^{A_2B_2}\frac{1}{4(p^0)^2}\nn
			\eea
	
		Instead of doing the time integrals first, lets instead do some of the integrals in $k_1,k_2$. Specifically, we will do the integrals over $k_1^+, k_2^+$ exploiting the fact that $D_>(k)$ is independent of $k^+$.  To that end, we can write the energy exponents in a manifestly covariant form and apply power counting of our Glauber modes, 
		\bea
		-k_1^0+q^0 -|\vec{q}-\vec{k}_1| = \frac{(q-k_1)^2}{-k_1^0+q^0 +|\vec{q}-\vec{k}_1|} \rightarrow (q^+-k_1^+)- \frac{(\vec{q}_{\perp}-\vec{k}_{1,\perp})^2}{q^-}\nn\\
		q^0-k_1^0-k_2^0-p^0= q^0-k_1^0-k_2^0-|\vec{q}-\vec{k}_1-\vec{k}_2| \rightarrow (q^+-k_1^+-k_2^+)- \frac{(\vec{q}_{\perp}-\vec{k}_{1,\perp}-\vec{k}_{2,\perp})^2}{q^-}\nn
		\eea
 Since p is on-shell and $p^0 =|\vec{p}|= |\vec{q}-\vec{k}_1-\vec{k}_2|$, we can write 
	\bea
	p^- = q^-, \ \ \vec{p}_{\perp}= \vec{q}_{\perp}-\vec{k}_{1,\perp}-\vec{k}_{2,\perp}, \ \ p^+ = \frac{p_{\perp}^2}{p^-}
	\eea
		so that it does not depend on $k_1^+, k_2^+$. 
 We are therefore free to do the integrals over $k_1^+,k_2^+$,
	
		\bea
		&&\int dk_1^+ dk_2^+\int_0^t dx_1^0 \int_0^t dx_2^0\int_0^t dy_1^0\int_0^t dy_2^0\Theta(x_1^0-x_2^0)\Theta(y_1^0-y_2^0)\nn\\
		&\times& e^{i(x_1^0-y_1^0+y_2^0-x_2^0) ((q^+-k_1^+)- \frac{(\vec{q}_{\perp}-\vec{k}_{1,\perp})^2}{q^-})} e^{i(x_2^0-y_2^0) ( (q^+-k_1^+-k_2^+)- \frac{(\vec{q}_{\perp}-\vec{k}_{1,\perp}-\vec{k}_{2,\perp})^2}{q^-})}\nn\\
		&=& (2\pi)^2\int dx_1^0 \int dx_2^0 \Theta(x_1^0-x_2^0)\nn\\
		&=& (2\pi)^2\frac{t^2}{2!}
		\eea

We can write $D_>^{AB}= D_>\delta^{AB}$, then do the trace and average over initial spin and color which gives us 
\bea
 	\Sigma^{(2)}&=& \frac{(|C_{qq}|^2C_FT_Ft)^2}{2!}\int \frac{d^2k_{1,\perp}}{(2\pi)^3}\frac{1}{(k_{1,\perp})^4}D_>(k_1) \int \frac{d^2k_{2,\perp}}{(2\pi)^3}\frac{1}{(k_{2,\perp})^4}D_>(k_2) \nn\\
		&\times&\langle q-k_1-k_2|\rho_n(0)|q-k_1-k_2\rangle  
\eea
When we go to impact parameter space, the convolution in $k_{i,\perp}$ turns into a product and this term becomes the second term in the exponential series, the first term given by Eq. \ref{Hg2R}. Given our discussion in Section \ref{sec:TwoG}, we see that this term will become relevant for $ t/t_I \geq 1$. If we keep computing this topology, i.e independent planar scattering at higher order in  $H_G$, then we see that the result in the regime $t/t_I \sim 1$ forms a series that sums into an exponent. This was also seen in \cite{Vaidya:2020cyi}, where only the information obtained form the O($H_G^2$) term was used to derive a differential equation for the time evolution of the reduced density matrix element of the collinear quark. The solution was found to be an exponent of the parameter t. 
 
We also have another diagram which has a non-planar topology as shown in Fig. \ref{Hg4np}. For this case, the collinear matrix element remains unchanged Eq. \ref{CollHG4}, but the Soft matrix element is given by Eq. \ref{HG4NP}.

\begin{figure}[h!]
   \begin{minipage}{.3\textwidth}
    \centering
    \includegraphics[width=0.9\textwidth]{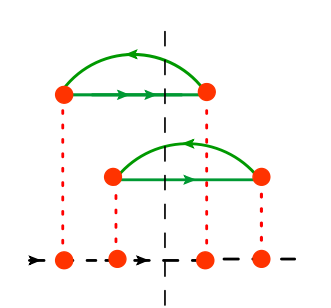}
    \caption{}
		\label{Hg4np}
  \end{minipage}
	 \begin{minipage}{.7\textwidth}
    \begin{align}
     S^{(4)}_{I}&=\langle X_s| \frac{1}{\mathcal{P}_{\perp}^2}O_s^{A_1}(x_1)\frac{1}{\mathcal{P}_{\perp}^2} O_s^{B_2}(y_2)\rho_B  |X_s\rangle  \nn\\
		&\langle X_s| \frac{1}{\mathcal{P}_{\perp}^2}O_s^{A_2}(x_2)\frac{1}{\mathcal{P}_{\perp}^2} O_s^{B_1}(y_1)\rho_B |X_s\rangle \nn\\
		\label{HG4NP}
		 \end{align}
		 \end{minipage}%
\end{figure}
		which immediately leads to 
		\bea
		S^{(4)}_{I}= \int \frac{d^2k_{1,\perp}dk_1^+}{(2\pi)^4}\frac{e^{-ik_1\cdot(x_1-y_2)}}{(k_{1,\perp})^4}D_>^{A_1B_2} \int \frac{d^2k_{2,\perp}dk_2^+}{(2\pi)^4}\frac{e^{-ik_2\cdot(x_2-y_1)}}{(k_{2,\perp})^4}D_>^{A_2B_1}
		\eea
		Now combining this with the collinear sector Eq.\ref{CollHG4} and following the same series of steps as for the planar diagram, 	
		\bea
		\Sigma_R^{(2)}&=&|C_{qq}|^2\int dx_1^0 \int dx_2^0 \int dy_1^0 \int dy_2^0\Theta(x_1^0-x_2^0) \Theta(y_1^0-y_2^0) \nn\\
		&\times& \frac{1}{(q^-)^2}\bar u(q)e^{iq^0 x_1^0}\frac{\slashed{\bar n}}{2}T^{A_1}e^{-i(|\vec{q}-\vec{k}_1|-i\Gamma_T)(x_1^0-x_2^0)}\frac{\slashed{ n}}{2}\frac{\slashed{\bar n}}{2}T^{A_2}u(p_1)e^{-ip^0 x_2^0}\langle p|\rho_n(0)|p\rangle \nn\\
	&\times&\bar u(p)e^{ip^0 y_2^0}\frac{\slashed{\bar n}}{2}T^{B_2}e^{i(|\vec{q}-\vec{k}_2|+i\Gamma_T)^0 (y_1^0-y_2^0)}\frac{\slashed{ n}}{2}\frac{\slashed{\bar n}}{2}T^{B_1}u(q)e^{-iq^0 y_1^0} \nn\\
	&\times& \int \frac{d^4k_1}{(2\pi)^4}\frac{e^{-ik_1 ^0 (x_1^0-y_2^0)}}{(k_{1,\perp})^4}D_>^{A_1B_2} \int \frac{d^4k_2}{(2\pi)^4}\frac{e^{-ik_2^0 (x_2^0-y_1^0)}}{(k_{2,\perp})^4}D_>^{A_2B_1}
		\eea
		
	Then applying power counting and performing the integrals over $k_1^+,k_2^+$, we get	
		\bea
		\delta(x_1^0-x_2^0+y_1^0-y_2^0) 
		\eea
		which sets the whole result to zero when combined with the time ordering constraints.
			This results also holds to higher orders at which then dictates that for independent scattering, the non-planar diagrams do not contribute at leading power in our EFT expansion parameters.
		
\item{Correlated scattering}\\
We now consider the case of correlated scattering which leads to a fully connected diagram as shown in Fig.\ref{Hg4c}. As before, the collinear matrix element remains unaffected so we focus on the soft matrix element. There are 4 possibilities here leading to a box or cross box diagram on each side of the cut. Here we will consider the box topology on either side of the cut to establish the relative importance of this diagram compared to the one for independent scattering. We will consider the more general case when we discuss the all order structure of the Glauber exchanges in the next section \ref{sec:AllOrder}.

\begin{figure}[h!]
   \begin{minipage}{.3\textwidth}
    \centering
    \includegraphics[width=0.9\textwidth]{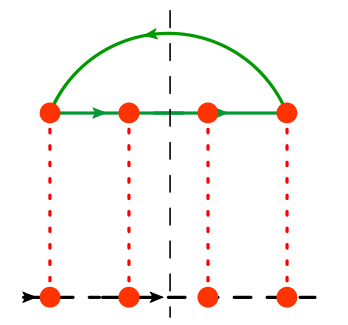}
    \caption{}
		\label{Hg4c}
  \end{minipage}
	 \begin{minipage}{.7\textwidth}
    \begin{align}
     S_c^{(4)}	&=\langle X_s| T\Big\{\frac{1}{\mathcal{P}_{\perp}^2}O_s^{A_1}(x_1)\frac{1}{\mathcal{P}_{\perp}^2} O_s^{A_2}(x_2)\Big\}\nn\\
		&\rho_B \bar T \Big\{\frac{1}{\mathcal{P}_{\perp}^2}O_s^{B_1}(y_1)\frac{1}{\mathcal{P}_{\perp}^2} O_s^{B_2}(y_2)\Big\} |X_s\rangle 
		 \end{align}
		 \end{minipage}%
\end{figure}

 		This lead to the fully connected result 
		\small
		\bea
		S_c^{(4)}&=&\int_0^t dx_1^0 \int_0^t dx_2^0 \int_0^t dy_1^0 \int_0^t dy_2^0 \int d^3x_1 \int d^3x_2 \int d^3y_1 \int d^3y_2\nn\\
		&\times&\int \frac{d^4p}{(2\pi)^4} \int \frac{d^4v_1}{(2\pi)^4} \int \frac{d^4v_2}{(2\pi)^4} \int \frac{d^4v_3}{(2\pi)^4}e^{-iv_1\cdot(x_1-x_2)}\slashed{v_1}\Big\{\frac{i}{v_1^2+i\epsilon}-2\pi \delta(v_1^2)n(|v_1^0|) \Big\}\nn\\
		&\times&e^{-iv_2\cdot(x_2-y_2)}\slashed{v_2}2\pi\Big\{\delta^+(v_2^2)- \delta(v_2^2)n(|v_2^0|) \Big\}e^{-iv_3\cdot(y_2-y_1)}\slashed{v_3}\Big\{\frac{-i}{v_3^2-i\epsilon}-2\pi \delta(v_3^2)n(|v_3^0|) \Big\} \nn\\
		&\times&e^{-ip\cdot(y_1-x_1)}\slashed{p}2\pi\Big\{\delta^+(-p^2)-\delta(p^2)n(|p^0|) \Big\}\frac{i}{(\vec{p}_{\perp}-\vec{v}_{1,\perp})^2}\frac{i}{(\vec{v}_{1,\perp}-\vec{v}_{2,\perp})^2}\frac{-i}{(\vec{v}_{2,\perp}-\vec{v}_{3,\perp})^2}\frac{-i}{(\vec{v}_{3,\perp}-\vec{p}_{\perp})^2}\nn
		\eea
		\normalsize
Combining with the collinear sector Eq.\ref{CollHG4}, and performing the co-ordinate integrals, we get 4 momentum conserving $\delta$ functions as before (We are taking the limit $t\rightarrow \infty$ here in anticipation that the final result will just be proportional to t)
\bea
 \delta^4(q-l_1-v_1+p)\delta^4(l_1+v_1-v_2-p_1)\delta^4(l_2-q+v_3-p) \delta^4(p_2-l_2+v_2-v_3) 
\eea
which once again leads to $p_1=p_2$, we can eliminate the integrals over  $p_1, p_2,l_1,l_2$ as before. At the same time, we define 
\bea
p-v_1= k_1, \ \ v_1-v_2= k_2, \ \ v_2-v_3=k_3,  
\eea
so that $p_1=p_2= q+k_1+k_2$. $l_1 =q+k_1 $, $l_2 =q+k_1+k_2+k_3$ ,
which finally yields for us the result 
\small
\bea
&&\Sigma_c^{(2)} = 4|C_{qq}|^4 t\int \frac{d^4p}{(2\pi)^4}\int \frac{d^4k_1}{(2\pi)^4}\frac{1}{(k_{1,\perp})^2}\int \frac{d^4k_2}{(2\pi)^4}\frac{1}{(k_{2,\perp})^2}\int \frac{d^4k_3}{(2\pi)^4}\frac{1}{(k_{3,\perp})^2}\frac{1}{(\vec{k}_{3,\perp}+\vec{k}_{2,\perp}+\vec{k}_{1,\perp})^2}  \nn\\
	&&\Big\{\text{Tr}\Big[T^{A_1}T^{B_1}T^{B_2}T^{A_2}\Big]\Big\}^2\langle q+k_1+k_2|\rho_n(0)|q+k_1+k_2\rangle \frac{iq^-}{(q+k_1)^2+i\epsilon}\frac{(-i)q^-}{(q+k_1+k_2+k_3)^2-i\epsilon} \nn\\
&\times& p^+\Big\{\frac{i}{(p-k_1)^2+i\epsilon}-2\pi \delta((p-k_1)^2)n(|(p-k_1)^0|) \Big\}\nn\\
&\times&2\pi p^+\Big\{\delta^+((p-k_1-k_2)^2)- \delta((p-k_1-k_2)^2)n(|(p-k_1-k_2)^0|) \Big\}\nn\\
&\times&p^+\Big\{\frac{-i}{(p-k_1-k_2-k_3)^2-i\epsilon}-2\pi \delta((p-k_1-k_2-k_3)^2)n(|(p-k_1-k_2-k_3)^0|) \Big\}\nn\\
&\times&2\pi p^+\Big\{\delta^+(-p^2)-\delta(p^2)n(|p^0|) \Big\}q^-\delta^+(q+k_1+k_2)^2
\eea
\normalsize
The result is a sum over several pieces (8 in all when we expand out all the brackets) and we can evaluate each term in turn. Our objective here is to determine a typical term from this diagram and see how it scales compared to the case of independent scatterers. Hence we consider one piece out of the 8 
\small
\bea
&& \widehat \Sigma^{(2)} = 4(2\pi)^2|C_{qq}|^4 t\int \frac{d^4p}{(2\pi)^4}\int \frac{d^4k_1}{(2\pi)^4}\frac{1}{(k_{1,\perp})^2}\int \frac{d^4k_2}{(2\pi)^4}\frac{1}{(k_{2,\perp})^2}\int \frac{d^4k_3}{(2\pi)^4}\frac{1}{(k_{3,\perp})^2}\frac{1}{(\vec{k}_{3,\perp}+\vec{k}_{2,\perp}+\vec{k}_{1,\perp})^2}  \nn\\
	&&\Big\{\text{Tr}\Big[T^{A_1}T^{B_1}T^{B_2}T^{A_2}\Big]\Big\}^2\langle q+k_1+k_2|\rho_n(0)|q+k_1+k_2\rangle \frac{iq^-}{(q+k_1)^2+i\epsilon}\frac{(-i)q^-}{(q+k_1+k_2+k_3)^2-i\epsilon}\nn\\
&\times& p^+\Big\{-2\pi \delta((p-k_1)^2)n(|(p-k_1)^0|)\Big\}p^+\Big\{\delta^+((p-k_1-k_2)^2)\Big\}\nn\\
&\times&p^+\Big\{\frac{-i}{(p-k_1-k_2-k_3)^2-i\epsilon} \Big\} p^+\Big\{-\delta(p^2)n(|p^0|) \Big\}q^-\delta^+(q+k_1+k_2)^2
\eea
\normalsize
Use the on-shell conditions to do the integral over $k_1^+$, $k_1^-$
 
\small
\bea
 &&\Sigma^{(2)} =  2|C_{qq}|^4t\int \frac{d^4p}{(2\pi)^4}\int \frac{d^2k_{1,\perp}}{(2\pi)^2}\frac{1}{(k_{1,\perp})^2}\int \frac{d^4k_2}{(2\pi)^4}\frac{1}{(k_{2,\perp})^2}\int \frac{d^4k_3}{(2\pi)^4}\frac{1}{(k_{3,\perp})^2}\frac{1}{(\vec{k}_{3,\perp}+\vec{k}_{2,\perp}+\vec{k}_{1,\perp})^2}  \nn\\
	&&\Big\{\text{Tr}\Big[T^{A_1}T^{B_1}T^{B_2}T^{A_2}\Big]\Big\}^2\langle q+k_1+k_2|\rho_n(0)|q+k_1+k_2\rangle \nn\\
	&\times&\frac{iq^-}{(-k_2^+)q^--(\vec{q}_{\perp}+\vec{k}_{1,\perp}+\vec{k}_{2,\perp})^2-(\vec{q}_{\perp}+\vec{k}_{1,\perp})^2+i\epsilon}\nn\\
	&\times& \frac{(-i)q^-}{k_3^+q^-+(\vec{q}_{\perp}+\vec{k}_{1,\perp}+\vec{k}_{2,\perp})^2-(\vec{q}_{\perp}+\vec{k}_{1,\perp}+\vec{k}_{2,\perp}+\vec{k}_{3,\perp})^2-i\epsilon}\nn\\
&\times& \Big\{-2\pi \text{sign}(p^+)n(|\frac{p^+}{2}+\frac{(\vec{p}_{\perp}-\vec{k}_{1,\perp})^2}{2p^+}|)\Big\}\Theta(p^+)\Big\{\delta^+(\frac{(\vec{p}_{\perp}-\vec{k}_{1,\perp})^2}{p^+}-k_2^--\frac{(\vec{p}_{\perp}-\vec{k}_{1,\perp}-\vec{k}_{2\perp})^2}{p^+})\Big\}\nn\\
&\times&\Big\{\frac{-i}{\frac{(\vec{p}_{\perp}-\vec{k}_{1,\perp})^2}{p^+}-k_2^--k_3^---\frac{(\vec{p}_{\perp}-\vec{k}_{1,\perp}-\vec{k}_{2\perp}-\vec{k}_{3,\perp})^2}{p^+}-i\epsilon/p^+} \Big\} p^+\Big\{-\delta(p^2)n(|p^0|) \Big\}
\eea
\normalsize
The integrals over $k_2^+,k_2^-,k_3^+,k_3^-$ contain rapidity divergence that are not regulated by dim. reg. To handle this, we therefore introduce rapidity regulators (\cite{Chiu:2012ir}) as shown in \cite{Rothstein:2016bsq} for $k_2^z$ and $k_3^z$ integrals and first do the integrals in $k_2^0$ using the on-shell condition and then $k_3^0$ using contour integration.
\small
\bea
 &&\Sigma^{(2)} = |C_{qq}|^4 t\int \frac{d^4p}{(2\pi)^4}\int \frac{d^2k_{1,\perp}}{(2\pi)^2}\frac{1}{(k_{1,\perp})^2}\int \frac{d^2k_{2,\perp}dk_2^z}{(2\pi)^4}\frac{1}{(k_{2,\perp})^2}\int \frac{d^2k_{3,\perp}dk_3^z}{(2\pi)^4}\frac{1}{(k_{3,\perp})^2}\frac{1}{(\vec{k}_{3,\perp}+\vec{k}_{2,\perp}+\vec{k}_{1,\perp})^2}  \nn\\
	&&\Big\{\text{Tr}\Big[T^{A_1}T^{B_1}T^{B_2}T^{A_2}\Big]\Big\}^2\langle q+k_1+k_2|\rho_n(0)|q+k_1+k_2\rangle \nn\\
	&\times&\frac{i|2k_2^z|^{-\eta}\nu^{\eta}}{2k_2^z+\tilde \Delta+i\epsilon}\frac{(-i)|2k_3^z|^{-\eta}\nu^{\eta}}{-2k_3^z+\Delta-i\epsilon}\Big\{-2\pi \text{sign}(p^+)n(|\frac{p^+}{2}+\frac{(\vec{p}_{\perp}-\vec{k}_{1,\perp})^2}{2p^+}|)\Big\}(-2\pi)\Theta(p^+) p^+\Big\{-\delta(p^2)n(|p^0|) \Big\}\nn
\eea
\normalsize 
where the precise for of $\Delta, \tilde \Delta $ is not important except that they are independent of $k_2^z,k_3^z$. 
Finally we can do the integrals over $k_2^z$ , $k_3^z$ which yield phase factors and make the full contribution real. 

\small
\bea
 &&\widehat \Sigma^{(2)} = (8\pi\alpha_s)^4 t\int \frac{d^2p_{\perp}\int_0^{\infty}dp^+}{(2\pi)^4}\int \frac{d^2k_{1,\perp}}{(2\pi)^2}\frac{1}{(k_{1,\perp})^2}\int \frac{d^2k_{2,\perp}}{(2\pi)^2}\frac{1}{(k_{2,\perp})^2}\nn\\
&\times&\int \frac{d^2k_{3,\perp}}{(2\pi)^2}\frac{1}{(k_{3,\perp})^2}\frac{1}{(\vec{k}_{3,\perp}+\vec{k}_{2,\perp}+\vec{k}_{1,\perp})^2} \Big\{\text{Tr}\Big[T^{A_1}T^{B_1}T^{B_2}T^{A_2}\Big]\Big\}^2\langle q+k_1+k_2|\rho_n(0)|q+k_1+k_2\rangle \nn\\
	&\times& n_F(|\frac{p^+}{2}+\frac{(\vec{p}_{\perp}-\vec{k}_{1,\perp})^2}{2p^+}|)n_F(|\frac{p^+}{2}+\frac{p_{\perp}^2}{2p^+}|)
	\eea
\normalsize 
Since this gives us a result proportional to t, we can combine it with the O($H_G^2$ ) term give a correction to the time scale $t_I$.
	
	\end{itemize}
	
If we apply the scaling behavior of the momenta involved, namely $k_{i,\perp} \sim T$ then we see that this term scales as $ t \alpha_s^4 T$, comparing this with the independent scattering piece which scales as $(t \alpha_s^2 T)^2$, we can draw the following conclusions 
\begin{itemize}
\item For $ t \alpha_s^2 T \sim 1$, with $\alpha_s <<1 $, only the independent scattering contribution is  important while the correlated scattering leads to a sub-leading contributions
\item 
For $\alpha_s <<1$, this particular correlated piece can give an O(1) contribution whenever $ t \sim 1/(T\alpha_s^4)$,  which then requires us to resum all contributions of the form $ (t T\alpha_s^4)^n$. This suggests that as t becomes larger, we need to include resummations of increasing number of coherent Glauber exchanges in order to maintain accuracy of our prediction.

\item Finally, the correlated scattering piece becomes as important as the independent scattering piece when $ \alpha_s^2 T \sim T$ , i.e. only when $\alpha_s \sim 1$, i.e., when we enter the non-perturbative regime. In that case we also have to resum all possible coherent Glauber exchanges, which we will consider in Section \ref{sec:AllOrder}.  We can also understand this intuitively as the regime where $t_I \sim t_e$, i.e., the correlation time in the medium becomes comparable to the interaction time with the system so that the environment retains information of the previous Glauber exchange with the medium which invalidates the Markovian approximation.  
\end{itemize}

In the next section, we will look at the structure of correlated scattering diagrams to all orders in perturbation theory. 

\section{System interaction time $t_I$ to all orders}
\label{sec:AllOrder}

In this section, we explore the structure of the system-bath interaction time to all orders in the Glauber Gluon exchange. From the analysis of previous section, we see that this amounts to to computing the complete set of diagrams with coherent Glauber exchanges between the system and the medium. We will first work out the amplitude for Soft-Collinear scattering and then look at its contribution to $t_I$. We follow the procedure in \cite{Rothstein:2016bsq}, making necessary changes to account for the thermal medium. 

\begin{itemize}

\item{$M^{(1)}$} 

\begin{figure}[h]
   \begin{minipage}{.3\textwidth}
    \centering
    \includegraphics[width=0.6\textwidth]{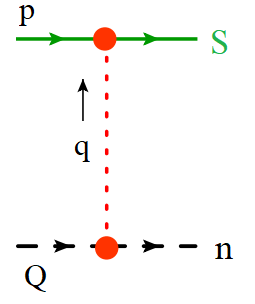}
    \caption{}
  \end{minipage}
	 \begin{minipage}{.5\textwidth}
    \begin{align}
        M^{(1)} &=  \frac{-2i g^2}{q_{\perp}^2} \bar u_s(p+q) \frac{\slashed{n}}{2}T^{A}u_s(p)\bar u_n(Q-q)\frac{\slashed{\bar n}}{2} T^{A} u_n(Q)\nn\\
				&\equiv \frac{-2i g^2}{q_{\perp}^2} T^A \otimes T^A S_{ns}
    \end{align}
		where we have defined 
		\begin{align}
		T^A \otimes T^A S_{ns} = \bar u_s(p+q) \frac{\slashed{n}}{2}T^{A}u_s(p)\bar u_n(Q-q)\frac{\slashed{\bar n}}{2} T^{A} u_n(Q)\nn
		\end{align} 
	  \end{minipage}%
\end{figure}

	which defines the action of the operator $T^A \otimes T^A$ on our Quark bilinear. We can generalize this to an operator definition 
	\small
\begin{align}
			T^{A_1}T^{A_2}...T^{A_N} \otimes T^{B_1}T^{B_2}...T^{B_M} S_{ns} = \bar u_s(p+q) \frac{\slashed{n}}{2}T^{A_1}T^{A_2}...T^{A_N}u_s(p)\bar u_n(Q-q)\frac{\slashed{\bar n}}{2} T^{B_1}T^{B_2}...T^{B_M}u_n(Q)\nn
\end{align} 
\normalsize
\item{$M^{(2)}$}

\begin{figure}[h!]
   \begin{minipage}{.25\textwidth}
    \centering
    \includegraphics[width=0.9\textwidth]{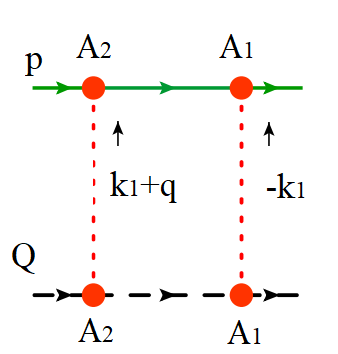}
    \caption{}
  \end{minipage}
	 \begin{minipage}{.75\textwidth}
    \small
		\begin{align}
      &= (-2ig^2)^2 T^{A_1}T^{A_2} \otimes T^{A_1}T^{A_2}S_{ns} \int \frac{d^4 k_1}{(2\pi)^4(\vec{k}_{1\perp}+\vec{q}_{\perp})^2}\frac{1}{\vec{k}_{1\perp}^2}\frac{iQ^-}{(Q-k_1-q)^2+i\epsilon}\nn\\
&\times \left(\frac{i}{(p+k_1+q)^2+i\epsilon}- 2\pi \delta((p+k_1+q)^2)n(|p^0+k_1^0+q^0|)\right)p^+\nn
    \end{align}
		\normalsize 
		 \end{minipage}%
\end{figure}
We can apply power counting for our Glauber mode and simplify this expression as 
\small
\begin{align}
&\mathcal{M}^{(2)} = (-2ig^2)^2 T^{A_1}T^{A_2} \otimes T^{A_1}T^{A_2}S_{ns} \int \frac{d^4 k_1}{(2\pi)^4(\vec{k}_{1\perp}+\vec{q}_{\perp})^2}\frac{1}{\vec{k}_{1\perp}^2}\frac{i}{Q^+-k_1^+-q^+- \frac{\vec{Q}_{\perp}-\vec{k}_{1,\perp}-\vec{q}_{\perp})^2}{Q^-}+\frac{i\epsilon}{Q^-}}\nn\\
&\Bigg\{\frac{i}{p^-+k_1^-+q^--\frac{(\vec{p}_{\perp}+\vec{k}_{1,\perp}+\vec{q}_{\perp})^2}{p^+}+\frac{i\epsilon}{p^+}}\nn\\
&- 2\pi \text{sign}(p^+)\delta\left(p^-+k_1^-+q^--\frac{(\vec{p}_{\perp}+\vec{k}_{1,\perp}+\vec{q}_{\perp})^2}{p^+}\right)n\left(|\frac{p^+}{2}+ \frac{p^++k_1^-+q^-}{2}|\right)\Bigg\}
\end{align}
\normalsize
We see that the integral over $k^+$ has a rapidity divergence which is not regulated by dim.reg and hence we need to put in rapidity regulator. We follow the procedure in \cite{Rothstein:2016bsq}. For each Glauber propagator with momentum l, we put in a factor $|2l^z|^{-\eta}\nu^{\eta}$ which then forces us to do the integral over energy using contours. We need to consider here that $Q^- >0$ while $p^+$ can be positive or negative since it is part of the thermal medium. We are then left with
\small
\bea
M^{(2)}&=& i(-2ig^2)^2 T^{A_1}T^{A_2}\otimes T^{A_1}T^{A_2}S_{ns}\int \frac{d^2 k_{1\perp}}{(2\pi)^2(\vec{k}_{1\perp}+\vec{q}_{\perp})^2}\frac{1}{k_{1\perp}^2}\nn\\
&\times&\left(\theta(p^+)- \text{sign}(p^+)n(|\frac{p^+}{2}+\frac{(\vec{p}_{\perp}+\vec{k}_{1\perp}+\vec{q}_{\perp})^2}{2p^+}|)\right)\int \frac{dk_1^z}{2\pi} \frac{|2k_1^z|^{-\eta}|2k_1^z|^{-\eta}\nu^{2\eta}}{2k_1^z +\Delta_1+i\epsilon}\nn\\
&=& 2\frac{(-ig^2)^2}{2!} T^{A_1}T^{A_2}\otimes T^{A_1}T^{A_2}S_{ns} \int \frac{d^2 k_{1\perp}}{(2\pi)^2(\vec{k}_{1\perp}+\vec{q}_{\perp})^2}\frac{1}{k_{1\perp}^2}\nn\\
&\times&\left(\theta(p^+)- \text{sign}(p^+)n(|\frac{p^+}{2}+\frac{(\vec{p}_{\perp}+\vec{k}_{1\perp}+\vec{q}_{\perp})^2}{2p^+}|)\right)\nn
\eea
\normalsize
where 
\begin{align}
\Delta_1=  Q^+-q^+-\frac{(\vec{Q}_{\perp}-\vec{k}_{1,\perp}-\vec{q}_{\perp})^2}{Q^-}+p^-+q^--\frac{(\vec{p}_{\perp}+\vec{k}_{1,\perp}+\vec{q}_{\perp})^2}{p^+}
\end{align}
and we have used the result 
\begin{align}
\int_{-\infty}^{\infty}\frac{dk^z}{2\pi}\frac{|2k^z|^{-2\eta}\nu^{2\eta}}{2k^z+\Delta+i\epsilon}= \frac{1}{4\pi}\Big[-i\pi+ \mathcal{O}(\eta)\Big]
\end{align}

Along with the box diagram, we also have a cross box diagram. Unlike the case of a vacuum background, where the cross box piece disappears, due to the presence of the thermal background, we also have a contribution from this piece. 

\begin{figure}[h!]
   \begin{minipage}{.3\textwidth}
    \centering
    \includegraphics[width=0.75\textwidth]{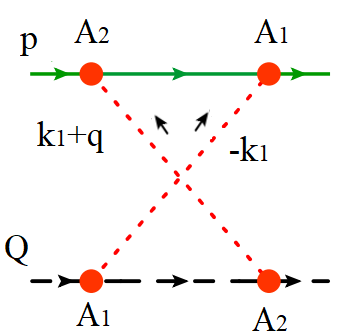}
    \caption{}
  \end{minipage}
	 \begin{minipage}{.7\textwidth}
	\small
    \begin{align}
      &= (-2ig^2)^2 T^{A_1}T^{A_2} \otimes T^{A_2}T^{A_1}S_{ns} \int \frac{d^4 k_1}{(2\pi)^4(\vec{k}_{1\perp}+\vec{q}_{\perp})^2}\frac{1}{\vec{k}_{1\perp}^2}\frac{iQ^-}{(Q+k_1)^2+i\epsilon}\nn\\
&\times \left(\frac{i}{(p+k_1+q)^2+i\epsilon}- 2\pi \delta((p+k_1+q)^2)n(|p^0+k_1^0+q^0|)\right)p^+\nn\\
    \end{align}
		\normalsize
		 \end{minipage}%
\end{figure}
which evaluates to 
\begin{align}
M^{(2)}_{\text{Cbox}}&=2\frac{(-ig^2)^2}{2!} T^{A_1}T^{A_2}\otimes T^{A_2}T^{A_1}S_{ns} \int \frac{d^2 k_{1\perp}}{(2\pi)^2(\vec{k}_{1\perp}+\vec{q}_{\perp})^2}\frac{1}{k_{1\perp}^2}\nn\\
&\times \left(-\theta(-p^+)- \text{sign}(p^+)n(|\frac{p^+}{2}+\frac{(\vec{p}_{\perp}+\vec{k}_{1\perp}+\vec{q}_{\perp})^2}{2p^+}|)\right)\nn 
\end{align}
We see that the only difference compared to the box diagram is the commutated color structure and the replacement $\theta(p^+) \rightarrow -\theta(-p^+)$. We will see that this simple substitution will also hold for $M^{(3)}$ and we can then generalize this to arbitrary order.

\item{$M^{(3)}$}\\
For 3 Glauber exchanges, we have 6 diagrams. Following the same steps as before, we can write the result  for each diagram  

\begin{figure}[h!]
   \begin{minipage}{.3\textwidth}
    \centering
    \includegraphics[width=0.9\textwidth]{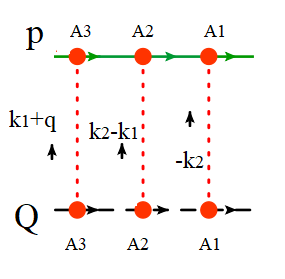}
    \caption{}
  \end{minipage}
	 \begin{minipage}{.5\textwidth}
    \small
		\begin{align}
        &=  \frac{-2 i^3g^6}{3!}  T^{A_1}T^{A_2}T^{A_3} \otimes T^{A_1}T^{A_2}T^{A_3}S_{ns} \int \frac{d^2 k_{2\perp}}{(2\pi)^2(\vec{k}_{1\perp}-\vec{k}_{2\perp})^2}\frac{1}{k_{2\perp}^2}\nn\\
&\times\int \frac{d^2 k_{1\perp}}{(2\pi)^2(\vec{k}_{1\perp}+\vec{q}_{\perp})^2}\frac{1}{k_{1\perp}^2}\left(\theta(p^+)- \text{sign}(p^+)n(|\frac{p^+}{2}+\frac{(\vec{p}_{\perp}+\vec{k}_{1\perp}+\vec{q}_{\perp})^2}{2p^+}|)\right)\nn\\
&\times\left(\theta(p^+)- \text{sign}(p^+)n(|\frac{p^+}{2}+\frac{(\vec{p}_{\perp}+\vec{k}_{2\perp}+\vec{q}_{\perp})^2}{2p^+}|)\right)\nn
    \end{align}
		\normalsize
		 \end{minipage}%
\end{figure}

\begin{figure}[h!]
   \begin{minipage}{.3\textwidth}
    \centering
    \includegraphics[width=0.9\textwidth]{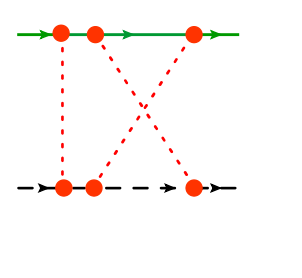}
    \caption{}
		\label{Mb3}
  \end{minipage}
	 \begin{minipage}{.5\textwidth}
    \small
		\begin{align}
       &=  \frac{-2i^3g^6}{3!} T^{A_1}T^{A_2}T^{A_3} \otimes T^{A_2}T^{A_1}T^{A_3}S_{ns} \int \frac{d^2 k_{2\perp}}{(2\pi)^2(\vec{k}_{1\perp}-\vec{k}_{2\perp})^2}\frac{1}{k_{2\perp}^2}\nn\\
&\times\int \frac{d^2 k_{1\perp}}{(2\pi)^2(\vec{k}_{1\perp}+\vec{q}_{\perp})^2}\frac{1}{k_{1\perp}^2}\left(\theta(p^+)- \text{sign}(p^+)n(|\frac{p^+}{2}+\frac{(\vec{p}_{\perp}+\vec{k}_{1\perp}+\vec{q}_{\perp})^2}{2p^+}|)\right)\nn\\
&\times\left(-\theta(-p^+)- \text{sign}(p^+)n(|\frac{p^+}{2}+\frac{(\vec{p}_{\perp}+\vec{k}_{2\perp}+\vec{q}_{\perp})^2}{2p^+}|)\right)
\label{nine}
    \end{align}
		
		\normalsize
		 \end{minipage}%
\end{figure}
The details for this result are presented in Appendix  \ref{app:rapid}

Similarly, we can work out the result for all the other topologies shown in Fig.\ref{3G}. 

\begin{figure}[h!]
  \includegraphics[width=\textwidth]{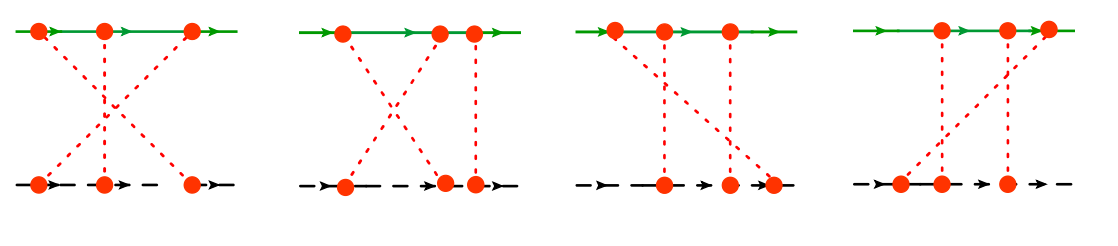}
  \caption{Diagrams for three Glauber exchanges.}
  \label{3G}
\end{figure}
The full result can then be expressed in a compact form.
\bea
 \Sigma^{(3)}&=&\frac{-2i^3g^6}{3!}\int d^2 k_{1\perp}\int d^2 k_{2\perp}\frac{1}{(2\pi)^2(\vec{k}_{1\perp}+\vec{q}_{\perp})^2}\frac{1}{(2\pi)^2k_{1\perp}^2}\frac{1}{(\vec{k}_{1\perp}-\vec{k}_{2\perp})^2}\nn\\
&\times& \sum_{\text{permutations}}F_3^{A_1A_2A_3}(k_{1,\perp},k_{2,\perp},q_{\perp})S_{ns}
\eea
where 
\bea
F_3^{A_1A_2A_3}(k_{1,\perp},k_{2,\perp},q_{\perp})&=&T^{A_1}T^{A_2}T^{A_3} \otimes T^{A_1}\left(\theta(p^+)- \text{Sign}(p^+)n(|\frac{p^+}{2}+\frac{(\vec{p}_{\perp}+\vec{k}_{1\perp}+\vec{q}_{\perp})^2}{2p^+}|)\right)T^{A_2}\nn\\
&\times&\left(\theta(p^+)- \text{Sign}(p^+)n(|\frac{p^+}{2}+\frac{(\vec{p}_{\perp}+\vec{k}_{2\perp}+\vec{q}_{\perp})^2}{2p^+}|)\right)T^{A_3} 
\eea
where we have written the term in a suggestive manner, namely each propagator is placed between color matrices $T_i, T_j$, and first term of the propagator can be set to either $\theta(p^+)$ if $ j >i$ or $-\theta(-p^+)$ otherwise. The sum over permutations (which are N! for N rungs) sums over all of the color matrix permutations. This means that 
\begin{align}
F_3^{A_1A_3A_2}&=T^{A_1}T^{A_2}T^{A_3} \otimes T^{A_1}\left(\theta(p^+)- \text{Sign}(p^+)n(|\frac{p^+}{2}+\frac{(\vec{p}_{\perp}+\vec{k}_{1\perp}+\vec{q}_{\perp})^2}{2p^+}|)\right)T^{A_3}\nn\\
&\times\left(-\theta(-p^+)- \text{Sign}(p^+)n(|\frac{p^+}{2}+\frac{(\vec{p}_{\perp}+\vec{k}_{2\perp}+\vec{q}_{\perp})^2}{2p^+}|)\right)T^{A_2}  
\end{align}
and so on. 

\item{$M^{(N)}$}

This then suggests a general rule for writing down the result for all possible ladder diagrams at order N 
\bea
 \Sigma^{(N)}&=& \frac{(-ig^2)^{N}}{N!} \int \frac{d^2 k_{1\perp}}{(2\pi)^2}\frac{1}{(\vec{k}_{1\perp}+\vec{q}_{\perp})^2}\Big\{\prod_{i=2}^{N-1}\int \frac{d^2 k_{i\perp}}{(2\pi)^2}\frac{1}{(\vec{k}_{i-1\perp}-\vec{k}_{i\perp})^2}\Big\}\frac{1}{k_{N-1,\perp}^2}\nn\\
&\times& \sum_{\text{permutations}}F_N^{A_1A_2...A_N}(k_{1,\perp},k_{2,\perp}..,k_{N-1,\perp},q_{\perp})2S_{ns}\nn\\
&=& \hat \Sigma^{(N)}2S_{ns}
\label{sig}
\eea
where we can think of $\hat \Sigma$ as an operator in color space acting on the collinear and soft currents.

\end{itemize} 

The complete amplitude is now given as 
\bea
\mathcal{M} = \sum_{i=1}^{\infty} \hat \Sigma^{(i)} 2S_{ns} 
\eea


\section{ Contribution to the reduced density matrix}
\label{sec:RDen}
We now want to evaluate the contribution of all the real diagrams to the evolution of the reduced density matrix. To do this, we need to square the amplitude obtained here and integrate over the loop momenta , while introducing the matrix element of the density matrix. The idea is that this should yield a redundant factor of t and therefore provide us with a term linear in t that will lead to a master equation. Following the same steps that led to Eq. \ref{Hg2R}   for the case of the square of a single Glauber exchange, the result for $\Sigma_R(T)$ defined in Eq.\ref{Den} is 

\bea
 \Sigma_R(t) &=&- \frac{8t}{N_c}\int \frac{d^4q}{(2\pi)^3} \int \frac{d^4 p}{(2\pi)^2}   \text{Tr}\Bigg[\Big\{\sum_{i=1}^{\infty} \hat \Sigma^{(i)}(\vec{p},\vec{q}_{\perp}) \Big\}\Big\{\sum_{i=1}^{\infty} \hat \Sigma^{(i)}(\vec{p},\vec{q}_{\perp}) \Big\}^{\dagger}\Bigg]\langle Q+q|\rho(0)|Q+q\rangle \nn\\
&\times&Q^-\delta^+((Q+q)^2)  p^+\left( \delta^+((-p)^2)-\delta(p^2)n\left(|p^0|\right)\right)p^+\left( \delta^+(p+q)^2-\delta((p+q)^2)n\left(|p^0+q^0|\right)\right)\nn
\eea

We can use the power counting for the q momentum (which is a Glauber mode) and perform the integrals over $q^+,q^-,p^-$.  
\bea
\Sigma_R(t)&=&-\frac{2t}{N_c}\int \frac{d^2q_{\perp}}{(2\pi)^3} \int \frac{d^2 p_{\perp}dp^+}{(2\pi)^2}   \text{Tr}\Bigg[\Big\{\sum_{i=1}^{\infty} \hat \Sigma^{(i)}(\vec{p},\vec{q}_{\perp}) \Big\}2\pi\left( \Theta(p^+)-\text{sign}(p^+)n\left(|\frac{p^+}{2}+\frac{(\vec{p}_{\perp}+\vec{q}_{\perp})^2}{2p^+}|\right)\right)\nn\\
&\times&\Big\{\sum_{i=1}^{\infty} \hat \Sigma^{(i)}(\vec{p},\vec{q}_{\perp}) \Big\}^{\dagger}\Bigg]\langle Q+q|\rho(0)|Q+q\rangle \left(-\Theta(-p^+)-\text{sign}(p^+)n\left(|\frac{p^+}{2}+\frac{p_{\perp}^2}{2p^+}|\right)\right)
\eea

We see that the total momentum transferred across the cut is $\vec{q}_{\perp}$, Having written the result in this form allows us to write the result as a convolution 
\bea
\Sigma_R(t) &=&t \int d^2k_{\perp} F(\vec{k}_{\perp}) \langle Q+k|\rho(0)|Q+k\rangle 
\eea
with 
\bea
 F(\vec{k}_{\perp})&=& -\frac{2}{N_c}\int \frac{d^2 p_{\perp}}{(2\pi)^5}dp^+   \text{Tr}\Bigg[\Big\{\sum_{i=1}^{\infty} \hat \Sigma^{(i)}(\vec{p},\vec{k}_{\perp}) \Big\}\left( \Theta(p^+)-\text{sign}(p^+)n\left(|\frac{p^+}{2}+\frac{(\vec{p}_{\perp}+\vec{k}_{\perp})^2}{2p^+}|\right)\right)\nn\\
&\times&\Big\{\sum_{i=1}^{\infty} \hat \Sigma^{(i)}(\vec{p},\vec{k}_{\perp}) \Big\}^{\dagger}\Bigg] \left(-\Theta(-p^+)-\text{sign}(p^+)n\left(|\frac{p^+}{2}+\frac{p_{\perp}^2}{2p^+}|\right)\right)
\eea

We can also readily write down the result for the virtual contribution to the density matrix evolution, which preserves the trace of the density matrix 
\bea
 \sigma_V(t)&=& -t \langle Q|\rho(0)|Q\rangle  \int d^2k_{\perp} F(\vec{k}_{\perp})
\eea

Therefore, for short times, we can write 
\bea
\langle Q |\rho(t) |Q \rangle =  (1-t) \langle Q|\rho(0)|Q\rangle  \int d^2k_{\perp} F(\vec{k}_{\perp})+ t \int d^2k_{\perp} F(\vec{k}_{\perp}) \langle Q+k|\rho(0)|Q+k\rangle 
\eea
In the limit $t \rightarrow 0$, this leads us to the master equation 
\bea
\partial_t P(Q^-,Q_{\perp},t) = - R P(Q^-,Q_{\perp},t)+   \int d^2k_{\perp} F(\vec{k}_{\perp}) P(Q^-,\vec{Q}_{\perp}+\vec{k}_{\perp},t)
\eea

with 
\bea
R = \int d^2k_{\perp} F(\vec{k}_{\perp}), \ \ \  P(Q^-,Q_{\perp},t)=\langle Q |\rho(t) |Q \rangle
\eea

This master equation can be solved as in our paper by moving to impact parameter space as shown in \cite{Vaidya:2020cyi} to yield 
\bea
P(Q^-,Q_{\perp},t) = \frac{f(Q^-)}{(2\pi)^2}\int d^2r_{\perp}e^{-i \vec{r}_{\perp}\cdot \vec{Q}_{\perp}}e^{\Big[-R(Q)+\tilde{F}(Q,-r_{\perp})\Big]t}
\eea
where $\tilde F$ is the function F in impact parameter space while $f(Q^-)/(2\pi)^2$ is the initial state density matrix element which we assume starts off with a quark withe zero transverse momentum. 
This final expression then resums, in principle $all$ the corrections, both those from independent as well as correlated diagrams. Of course, the problem now reduces to computing $F(k_{\perp})$ either analytically or numerically by performing all the transverse momentum integrals in the operator $\hat \Sigma$ defined order by order in Eq. \ref{sig}. 
I now turn to the evaluation of the F($k_{\perp}$) kernel.


\section{Simplification for the kernel}

\label{sec:Simp}

We now look at possible resummation/ simplification for the kernel $F(\vec{k}_{\perp})$. 
Lets look a the expression for $F(k_{\perp})$ again 
\bea
  F(\vec{k}_{\perp})&=& -\frac{2}{N_c}\int \frac{d^2 p_{\perp}}{(2\pi)^5}dp^+   \text{Tr}\Bigg[\Big\{\sum_{i=1}^{\infty} \hat \Sigma^{(i)}(\vec{p},\vec{k}_{\perp}) \Big\}\left( \Theta(p^+)-\text{sign}(p^+)n\left(|\frac{p^+}{2}+\frac{(\vec{p}_{\perp}+\vec{k}_{\perp})^2}{2p^+}|\right)\right)\nn\\
&\times&\Big\{\sum_{i=1}^{\infty} \hat \Sigma^{(i)}(\vec{p},\vec{k}_{\perp}) \Big\}^{\dagger}\Bigg]\left(-\Theta(-p^+)-\text{sign}(p^+)n\left(|\frac{p^+}{2}+\frac{p_{\perp}^2}{2p^+}|\right)\right)\nn\\
&=& \frac{2}{N_c}\int \frac{d^2 p_{\perp}}{(2\pi)^5}dp^+   \text{Tr}\Bigg[\Big\{\sum_{i=1}^{\infty} \hat \Sigma^{(i)}(\vec{p},\vec{k}_{\perp}) \Big\}\Big\{\sum_{i=1}^{\infty} \hat \Sigma^{(i)}(\vec{p},\vec{k}_{\perp}) \Big\}^{\dagger}\Bigg]\nn\\
&\times& \Bigg\{\Theta(p^+)n_F\left(|\frac{p^+}{2}+\frac{p_{\perp}^2}{2p^+}|\right)\left(1-n_F\left(|\frac{p^+}{2}+\frac{(\vec{p}_{\perp}+\vec{k}_{\perp})^2}{2p^+}|\right)\right)\nn\\
&+&\Theta(-p^+)n_F\left(|\frac{p^+}{2}+\frac{(\vec{p}_{\perp}+\vec{k}_{\perp})^2}{2p^+}|\right)\left(1-n_F\left(|\frac{p^+}{2}+\frac{p_{\perp}^2}{2p^+}|\right)\right)\Bigg\}
\label{FSimp}
\eea

Due to the factorial growth of the color structures, in general it will not be possible to do an all order resummation for $F(k_{\perp})$. We can however, look at certain limits in which the analysis is more tractable. 

\subsection{$k_{\perp} >> T$}

One limit we can consider is  $k_{\perp} >> T $. If we do this, then the only pieces that remain in the $\Sigma$ terms are those which have either all terms proportional to $\Theta(p^+)$ or those with $\Theta(-p^+)$. These correspond to the diagrams which either have all rungs in order or all of them inverted. At the same time looking at Eq. \ref{FSimp}, we see that that $\Theta(-p^+)$ term reduces to 0 so that only the $\Theta(p^+)$ channel contributes. The operator $\hat \Sigma$ defined in Eq.\ref{sig} now becomes 

\bea
\hat \Sigma^{(N)} &=& \sum_{\text{permutations}}F_N^{T^{A_1}T^{A_2}..T^{A_N}}(k_{1,\perp},k_{2,\perp}..,k_{N-1,\perp},q_{\perp})\xrightarrow{k_{i,\perp}>>T}\nn\\
&&\hat \Sigma^{(N)}\Big|_{k_{\perp}>>T} =\frac{(-ig^2)^N}{N!}\int \frac{d^2 k_{1\perp}}{(2\pi)^2}\frac{1}{(\vec{k}_{1\perp}+\vec{q}_{\perp})^2}\Big\{\prod_{i=2}^{N-1}\int \frac{d^2 k_{i\perp}}{(2\pi)^2}\frac{1}{(\vec{k}_{i-1\perp}-\vec{k}_{i\perp})^2}\Big\}\frac{1}{k_{N-1,\perp}^2}\nn\\
&\times& \Bigg\{\Theta(p^+) T^{A_1}...T^{A_N}\otimes T^{A_1}...T^{A_N}\Bigg\} \nn\\
&\equiv& \frac{(-ig^2)^N}{N!}\int \frac{d^2 k_{1\perp}}{(2\pi)^2}\frac{1}{(\vec{k}_{1\perp}+\vec{q}_{\perp})^2}\Big\{\prod_{i=2}^{N-1}\int \frac{d^2 k_{i\perp}}{(2\pi)^2}\frac{1}{(\vec{k}_{i-1\perp}-\vec{k}_{i\perp})^2}\Big\}\frac{1}{k_{N-1,\perp}^2}\Theta(p^+)\hat c^N \nn\\
\eea
where we have defined the operator 
\bea
\hat c =  T^A \otimes T^A,  \ \ \ \text{such that} \nn\\
(\hat c)^N = T^{A_1}...T^{A_N}\otimes T^{A_1}...T^{A_N}
\label{OpC}
\eea

We can simplify the form of this expression considerably by writing out the result in impact parameter space 
\bea
 \hat \Sigma^{(N)}(b)\Big|_{k_{\perp}>>T} &=&\int \frac{d^2\vec{q}_{\perp}}{(2\pi)^2}e^{-i\vec{q}_{\perp}\cdot \vec{b}} \hat \Sigma^{(N)}(b)\Big|_{k_{\perp}>>T}
\equiv \frac{(i\phi(b))^N}{N!}
\eea
where 
\bea
\phi(b) =  (-ig^2)\Theta(p^+)\hat c \int \frac{d^2k_{\perp}}{(2\pi)^2} \frac{e^{-i \vec{k}_{\perp}\cdot \vec{b}}}{k_{\perp}^2} 
\eea
which now allows us to write our all orders result as 
\bea
\sum_{i=1}^{\infty}  \hat \Sigma^{(i)}\Big|_{k_{\perp}>>T} =  \int d^2b_Le^{i\vec{k}_{\perp} \cdot \vec{b}} \Big\{e^{i\phi(b)}-1 \Big\}
\eea

We can now look at our kernel
\small 
\bea
F(k_{\perp})\Big|_{k_{\perp}>>T}&=& \frac{2}{N_c}\int \frac{d^2 p_{\perp}}{(2\pi)^5}dp^+  \int d^2b_Le^{i\vec{k}_{\perp} \cdot \vec{b}_L} \int d^2b_Re^{-i\vec{k}_{\perp} \cdot \vec{b}_R}  \text{Tr}\Bigg[\Big\{e^{i\phi(b_L)}-1 \Big\}\Big\{e^{i\phi(b_R)}-1 \Big\}^{\dagger}\Bigg]\nn\\
&\times& \Bigg\{\Theta(p^+)n_F\left(|\frac{p^+}{2}+\frac{p_{\perp}^2}{2p^+}|\right)\Bigg\}
\eea
\normalsize 
This allows us to do the integral over $p_{\perp},p+$ analytically 
\bea
\int p_{\perp}dp_{\perp}dp^+ \Theta(p^+)n_F\left(|\frac{p^+}{2}+\frac{p_{\perp}^2}{2p^+}|\right)= 3\zeta(3)
\eea
At the same time, we can do the integral over $b_L, b_R$ analytically, following \cite{Rothstein:2016bsq} , introducing a gluon mass $m_D$ to regulate the IR divergence in the $k_{i,\perp}$ integrals,

\bea
  \int d^2be^{-i\vec{k}_{\perp} \cdot \vec{b}}\Big\{ e^{i\phi(b)}-1\Big\}&=& \frac{i4\pi \hat c \alpha_s(\mu)}{-k_{\perp}^2}\frac{\Gamma[1+i\hat c \alpha_s(\mu)]}{\Gamma[1-i\hat c\alpha_s(\mu)]}\left(\frac{k_{\perp}^2}{m_D^2e^{2\gamma_E}}\right)^{-i \hat c \alpha_s(\mu)}\nn\\
	&=& \frac{i4\pi \hat c \alpha_s(\mu)}{-k_{\perp}^2}e^{i\delta(-k_{\perp}^2,\alpha_s)}
	\label{Int}
\eea
with 
\bea
\delta(-k_{\perp}^2,s) &=& -\hat c \alpha_s(\mu) \ln \left(\frac{q_{\perp}^2}{m_D^2}\right)+ 2\sum_{k=1}^{\infty} \frac{(-1)^{k+1}\zeta_{2k+1}}{2k+1}(\hat c \alpha_s(\mu))^{2k+1} 
\eea

When we combine this with the complex conjugate piece and perform the color trace, the phase factor $\delta$ cancels out completely leaving behind only a contribution from the leading order diagram. This is to be expected since in this limit we are throwing away the thermal contribution from all higher diagrams, with only vacuum contributions remaining. These corrections are known to resum into a phase factor which is what we find here. 

Therefore we finally have 
\bea
F(k_{\perp}) = \frac{2}{N_c}\frac{(4\pi)^2\alpha_s^2}{k_{\perp}^4}\text{Tr}\Big[T^A T^B\Big]\text{Tr}\Big[T^AT^B\Big]\frac{3 \zeta(3)}{(2\pi)^4}=\frac{6\alpha_s^2C_FT_FT^3}{\pi^2k_{\perp}^4} \zeta(3)
\eea
which is correct to all orders in perturbation theory.


\subsection{$k_{\perp} << T$}

The other interesting limit is when $ k_{\perp}<<T$. In that case, we can drop the $k_{\perp}$ dependence from all the Fermi distribution factors, which allows us to write 
\bea
 &&n\left(|\frac{p^+}{2}+\frac{(\vec{p}_{\perp}+\vec{q}_{\perp})^2}{2p^+}|\right)\rightarrow n\left(|\frac{p^+}{2}+\frac{\vec{p}_{\perp}^2}{2p^+}|\right)\nn\\
\eea
Then we can simplify our kernel in Eq. \ref{FSimp} as 
\bea
 F(k_{\perp})\Big|_{k_{\perp}<<T}&=& \frac{2}{N_c}\int \frac{d^2 p_{\perp}}{(2\pi)^5}dp^+   \text{Tr}\Bigg[\Big\{\sum_{i=1}^{\infty} \tilde \Sigma^{(i)}(\vec{p},\vec{k}_{\perp}) \Big\}\Big\{\sum_{i=1}^{\infty} \tilde \Sigma^{(i)}(\vec{p},\vec{k}_{\perp}) \Big\}^{\dagger}\Bigg]\nn\\
&\times& \Bigg\{n_F\left(|\frac{p^+}{2}+\frac{p_{\perp}^2}{2p^+}|\right)\left(1-n_F\left(|\frac{p^+}{2}+\frac{(\vec{p}_{\perp})^2}{2p^+}|\right)\right)\Bigg\}\nn\\
\eea
with 
\bea
 \tilde \Sigma^{(i)}(\vec{p},\vec{k}_{\perp})&=&  \frac{(-ig^2)^N}{N!}\int \frac{d^2 k_{1\perp}}{(2\pi)^2}\frac{1}{(\vec{k}_{1\perp}+\vec{q}_{\perp})^2}\Big\{\prod_{i=2}^{N-1}\int \frac{d^2 k_{i\perp}}{(2\pi)^2}\frac{1}{(\vec{k}_{i-1\perp}-\vec{k}_{i\perp})^2}\Big\}\frac{1}{k_{N-1,\perp}^2}\nn\\
&\times& \sum_{\text{permutations}}F_N^{A_1A_2..A_N}(p_{\perp},p^+)
\eea
where we have dropped all $k_{i,\perp}$ dependence from the factor $F_N$. This allows us to simplify our result by moving to impact parameter space, 
\bea
 \tilde \Sigma^{(N)}(\vec{p},\vec{b})&=& \frac{(i\Phi(b))^N}{N!}\sum_{\text{permutations}}F_N^{T^{A_1}T^{A_2}..T^{A_N}}(p_{\perp},p^+)
\eea

\bea
\Phi(b) =   -g^2\int \frac{d^2k_{\perp}}{(2\pi)^2}\frac{e^{-i \vec{k}_{\perp}\cdot \vec{b}}}{k_{\perp}^2} 
\eea

This object is not resummed easily to all in terms of a simple closed form at the amplitude level due to the factorial growth of the color structures. To proceed further, we than have to look at the amplitude squared and color traced structure order by order.

However, we can see that there are certain classes of diagrams which do cancel to all orders, when combined with their complex conjugate. If we look at the planar diagrams, which give the regular ladder diagrams to all order then we can write our result at the amplitude level in closed form 

 \bea
 F^{\text{ladder}}(k_{\perp})&=& \frac{2}{N_c}\int \frac{d^2 p_{\perp}}{(2\pi)^5}dp^+  \frac{\Bigg\{n_F\left(|\frac{p^+}{2}+\frac{p_{\perp}^2}{2p^+}|\right)\left(1-n_F\left(|\frac{p^+}{2}+\frac{(\vec{p}_{\perp})^2}{2p^+}|\right)\right)\Bigg\}}{\Big\{\theta(p^+)-\text{sign}(p^+)n\left(|\frac{p^+}{2}+\frac{p_{\perp}^2}{2p^+}|\right)\Big\}^2} \nn\\
&\times&  \int d^2b_Le^{i\vec{k}_{\perp} \cdot \vec{b}_L} \int d^2b_Re^{-i\vec{k}_{\perp} \cdot \vec{b}_R} \text{Tr}\Bigg[\Big\{e^{i\tilde \phi(b_L,p)}-1\Big\}\Big\{e^{-i \tilde \phi(b_R,p)}-1\Big\}\Bigg]\nn\\
\eea
with 
\bea
\tilde \phi(b,p) =   -g^2\Big\{\theta(p^+)-\text{sign}(p^+)n\left(|\frac{p^+}{2}+\frac{p_{\perp}^2}{2p^+}|\right)\Big\}\hat c \int \frac{ d^2k_{\perp}}{(2\pi)^2}\frac{e^{-i \vec{k}_{\perp}\cdot \vec{b}}}{k_{\perp}^2} 
\eea
with $\hat c$ being the same operator defined in Eq. \ref{OpC}. Therefore, using the same series of steps as Eq.\ref{Int}, we see that an all order cancellation goes through leaving behind only the tree level result. \\
We can also play the same game with the fully inverted ladder diagrams to all order. In this case we are counting the tree level result again so we can subtract it off to avoid double counting. We can therefore write 
\bea
 F^{\text{InvLadder}}(k_{\perp})&=& \frac{2}{N_c}\int \frac{d^2 p_{\perp}}{(2\pi)^5}dp^+   \frac{\Bigg\{n_F\left(|\frac{p^+}{2}+\frac{p_{\perp}^2}{2p^+}|\right)\left(1-n_F\left(|\frac{p^+}{2}+\frac{(\vec{p}_{\perp})^2}{2p^+}|\right)\right)\Bigg\}}{\Big\{-\theta(-p^+)-\text{sign}(p^+)n\left(|\frac{p^+}{2}+\frac{p_{\perp}^2}{2p^+}|\right)\Big\}^2} \nn\\
&\times& \int d^2b_Le^{i\vec{k}_{\perp} \cdot \vec{b}_L} \int d^2b_Re^{-i\vec{k}_{\perp} \cdot \vec{b}_R}\text{Tr}\Bigg[\Big\{e^{i \phi_I(b_L,p)}-1\Big\}\Big\{e^{-i\phi_I(b_R,p)}-1\Big\}\Bigg]\nn
\eea
where we have defined 
\bea
\phi_I(b,p) = -g^2\Big\{-\theta(-p^+)-\text{sign}(p^+)n\left(|\frac{p^+}{2}+\frac{p_{\perp}^2}{2p^+}|\right)\Big\}\hat d \int \frac{ d^2k_{\perp}}{(2\pi)^2}\frac{e^{-i \vec{k}_{\perp}\cdot \vec{b}}}{k_{\perp}^2} 
\eea
The operator $\hat d$ acts in an inverse manner compared to $\hat c$ such that 
\bea
(\hat d)^N =  T^{A_1}T^{A_2}...T^{A_N}\otimes T^{A_N}T^{A_{N_1}}....T^{A1} 
\eea

Naturally this begs the question if the hybrid diagrams, i.e., the ones which have some rungs in order and some others crossed, also exponentiate. We can already see that this is not possible since at the amplitude level,  the number of diagrams at each order grows factorially and not exponentially, so that all possible topologies cannot be recovered by simple exponentiation. 

However, a partial exponentiation still goes through which leads to some cancellation and hence simplifies the analysis at each order in $
\alpha_s$. We can see this by combining the ladder and inverse ladder into a single exponent, which will therefore also generate some of the hybrid diagrams.
\bea
  \widehat F(k_{\perp})&=& \frac{2}{N_c}\int \frac{d^2 p_{\perp}}{(2\pi)^5}dp^+  \frac{\Bigg\{n_F\left(|\frac{p^+}{2}+\frac{p_{\perp}^2}{2p^+}|\right)\left(1-n_F\left(|\frac{p^+}{2}+\frac{(\vec{p}_{\perp})^2}{2p^+}|\right)\right)\Bigg\}}{\Big\{\text{sign}(p^+)(1-2n\left(|\frac{p^+}{2}+\frac{p_{\perp}^2}{2p^+}|\right))\Big\}^2}\nn\\
&\times&\int d^2b_Le^{i\vec{k}_{\perp} \cdot \vec{b}_L} \int d^2b_Re^{-i\vec{k}_{\perp} \cdot \vec{b}_R}\text{Tr}\Bigg[\Big\{e^{i\tilde \phi(b_L,p)+i\phi_I(b_L,p)}-1\Big\}\Big\{e^{-i\phi(b_R,p)-i\phi_I(b_R,p)}-1\Big\}\Bigg]\nn
\eea
The action of the operators in the exponent, when expanded out, can be defined as 
\bea
(\hat d)^m(\hat c)^n =  T^{A_1}...T^{A_n}T^{A_{n+1}}...T^{A_{n+m}}\otimes T^{A_{n+m}} .... T^{A_{n+1}}T^{A_1}T^{A_2}...T^{A_n} 
\eea. 
There are however hybrid diagrams which remain uncanceled at each order and hence contribute to the kernel.\\
We see immediately that all the terms with odd powers of $\alpha_s$ cancel out with their complex conjugates. We therefore only have to consider even powers of $\alpha_s$.
The tree level result ($O(\alpha_s^2)$)  gives 
\begin{align}
F^{(1)}(k_{\perp}) &=\frac{2}{N_c}\text{Tr} \Big[T^{A_1}T^{A_2}\Big]\text{Tr}\Bigg [T^{A_1}T^{A_2}\Bigg]\int d^2b_L e^{i \vec{k}_{\perp} \cdot \vec{b}_L}\int d^2b_Re^{-i\vec{k}_{\perp} \cdot \vec{b}_R} i\phi(b_L)(-i\phi(b_R))\nn\\
&\times \int \frac{d^2 p_{\perp}}{(2\pi)^5}dp^+  \Big\{n_F\left(|\frac{p^+}{2}+\frac{p_{\perp}^2}{2p^+}|\right)\Big\}\Big\{1-n_F\left(|\frac{p^+}{2}+\frac{p_{\perp}^2}{2p^+}|\right)\Big\}\nn
\end{align}

with 
\bea
\phi(b) = -g^2 \int \frac{d^k_{\perp}}{(2\pi)^2} \frac{e^{-i\vec{k}_{\perp} \cdot \vec{b}}}{k_{\perp}^2+m_D^2} 
\eea
We see that the integral over p is decoupled from the $k_{\perp}$ dependence, We can do the integrals over $b_L$, $b_R$ by using the result from Eq.\ref{Int} which we rewrite here for convenience and then expand out order by order in $\phi(b)$, 
\bea
 \int d^2be^{-i\vec{k}_{\perp} \cdot \vec{b}}\Big\{ e^{i\phi(b)}-1\Big\}&=& \frac{i4\pi \alpha_s(\mu)}{-k_{\perp}^2}e^{i\delta(-k_{\perp}^2,\alpha_s)}
\eea
with 
\bea
\delta(t,s) &=& -\alpha_s(\mu) \ln \left(\frac{-t}{m^2}\right)+ 2\sum_{k=1}^{\infty} \frac{(-1)^{k+1}\zeta_{2k+1}}{2k+1}(\hat c \alpha_s(\mu))^{2k+1} 
\eea
Expanding out, we see that 
\bea
 \int d^2be^{-i\vec{k}_{\perp} \cdot \vec{b}}i\phi(b)= -\frac{i 4\pi \alpha_s}{k_{\perp}^2}, \ \ \  \ \int  d^2be^{-i\vec{k}_{\perp} \cdot \vec{b}}\frac{(i\phi(b)^3}{3!} = -\frac{(i)^34\pi\alpha_s^3}{2!k_{\perp}^2}\ln^2 \left(\frac{k_{\perp}^2}{m_D^2}\right) 
\eea
We can do the integral over $p$ exactly

\begin{align}
 \int d^2 p_{\perp}dp^+ \Big\{n_F\left(|\frac{p^+}{2}+\frac{p_{\perp}^2}{2p^+}|\right)\Big\}\Big\{1-n_F\left(|\frac{p^+}{2}+\frac{p_{\perp}^2}{2p^+}|\right)\Big\} = T^3 4\pi \frac{\pi^2}{3}
\end{align}

so that 
\begin{align}
F^{(1)}(k_{\perp})&= 2C_FT_F\frac{(4\pi\alpha_s)^2}{(2\pi)^5k_{\perp}^4}T^3 4\pi \frac{\pi^2}{3}= \frac{4C_FT_F\alpha_s^2T^3}{3k_{\perp}^4}
\end{align}

 We will also consider here the first non-trivial correction beyond the tree level diagram which goes as $\alpha_s^4$. We see that due to the partial exponentiation of terms, only two diagrams (and their c.c) remain uncanceled at this order. 

\small
\bea
F^{(2)}(k_{\perp})&=&\frac{8}{N_c}\text{Tr} \Big [T^{A_1}T^{A_2}T^{A_3}T^{A_4}\Big]\text{Tr}\Big [T^{A_2}T^{A_1}T^{A_3}T^{A_4}\Big]\int d^2b_L e^{i \vec{k}_{\perp} \cdot \vec{b}_L}\int d^2b_Re^{-i\vec{k}_{\perp} \cdot \vec{b}_R} \frac{(i\phi(b_L))^3}{3!}(-i\phi(b_R))\nn\\
&\times& \int \frac{d^2 p_{\perp}}{(2\pi)^5}dp^+ \Big\{\theta(p^+)-\text{sign}(p^+)n\left(|\frac{p^+}{2}+\frac{p_{\perp}^2}{2p^+}|\right)\Big\}^2\Big\{-\theta(-p^+)-\text{sign}(p^+)n\left(|\frac{p^+}{2}+\frac{p_{\perp}^2}{2p^+}|\right)\Big\}^2\nn\\
\eea
\normalsize 

 Again, we can do the integral over $p^+,p_{\perp}$ exactly, 
\bea
  \int d^2 p_{\perp}dp^+ \Big\{n_F\left(|\frac{p^+}{2}+\frac{p_{\perp}^2}{2p^+}|\right)\Big\}^2\Big\{1-n_F\left(|\frac{p^+}{2}+\frac{p_{\perp}^2}{2p^+}|\right)\Big\}^2 = 2\pi T^3\left(-\frac{2}{3}+\frac{\pi^2}{9}\right)
\eea
We can therefore write our result at O($\alpha_s^4$) as 
\begin{align}
F^{(1)}(k_{\perp})&= -\frac{8}{N_c}\Bigg\{\left(\frac{N_c^2-4}{N_c}\right)^2-N_c^2+\frac{C_F^2}{4}\Bigg\}\frac{T^3}{(2\pi)^4}\left(-\frac{2}{3}+\frac{\pi^2}{9}\right)\frac{(4\pi)^2\alpha_s^4}{2!k_{\perp}^4} \ln^2 \left(\frac{k_{\perp}^2}{m_D^2}\right)
\end{align}

This result is sensitive to the IR scale $m_D \sim gT$. Since we have taken the limit $k_{\perp} <<T$, we can expect that the logarithm here is small and does not lead to large corrections.  We can continue in this fashion and analytical results can be obtained at each order in perturbation theory.

The most general case when $k_{\perp} \sim T$ is not easily simplified beyond what I have already done, which is to provide all order integrands in transverse momentum space as defined in Eq. \ref{FSimp}. We then need to resort to numerical integration order by order in $\alpha_s^{2n}$ to make progress.

\section{Conclusion} 
\label{sec:Con}

In this paper, I have presented the results for the forward scattering of an energetic particle in a plasma at thermal equilibrium. I examine the radiative corrections arising from Glauber exchange diagrams between the jet and the medium to all orders in perturbation theory. 
I observe that there are three relevant time scales which describe the system, namely the medium correlation time, the system-medium interaction time and the time of propagation of the jet in the thermal medium. Depending on the hierarchy between these scales, different classes of diagrams become relevant at  each order in perturbation theory. All diagrams that give a coherent Glauber exchange can be collected into a kernel which contributes to the jet-medium interaction time $t_I$ . When the time of propagation of the jet in the medium, t becomes comparable to $t_I$, then a resummation of all terms proportional to $t/t_I$ can be carried out in the form of an exponent in impact parameter space. 

Using the effective field theory for forward scattering, which is formulated in terms of Glauber momentum modes mediating interactions between a soft medium and a collinear jet, I provide an expression for $t_I$ to all orders in Glauber exchange diagrams. Due to the presence of the thermal medium, there is, in general a factorial growth in the number of topologies which contribute at each order, all of which can be simplified as integrals over transverse momentum exchanged between the jet and the medium weighted by Fermi distribution functions (for a fermionic thermal bath). The measurement imposed on the final state is the transverse momentum broadening  $q_T$ of the energetic parton. An all order analytic expression for $t_I$ is obtained in the limit $q_T>>T$, in which case radiative corrections beyond leading order sums to a Glauber phase factor that cancels out with its complex conjugate. The other limit $q_T<<T$ also yields a simplified expression for $t_I$, which, while is not resummed, can be evaluated analytically order by order in perturbation theory.

For now, I have ignored any Soft and Collinear corrections, which is something I will include in a future work. The larger goal, is to systematically incorporate these corrections into the Effective Field Theory for jet substructure developed in \cite{Vaidya:2020lih}. 

\acknowledgments
I thank Xiaohui Liu for useful discussions during the initial stages of this project. This work is supported by the Office of Nuclear Physics of the U.S. Department of Energy under Contract DE-SC0011090 and Department of Physics, Massachusetts Institute of Technology.



\appendix

\section{ Rapidity integrals for Glauber exchange diagrams}
\label{app:rapid}
Here I derive the result in Eq. \ref{nine} corresponding to the Fig. \ref{Mb3}. 
The amplitude looks like 

\begin{figure}[h!]
   \begin{minipage}{.3\textwidth}
    \centering
    \includegraphics[width=\textwidth]{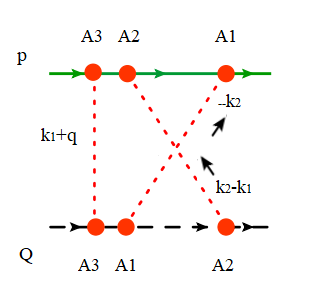}
    \caption{}
		\label{M3c}
  \end{minipage}
	 \begin{minipage}{.7\textwidth}
    \small
		\begin{align}
       &= (-2ig^2)^3T^{A_1}T^{A_2}T^{A_3} \otimes T^{A_2}T^{A_1}T^{A_3} S_{ns}\nu^{3\eta} \int \frac{d^4k_1|2k_1^z|^{-\eta}}{(2\pi)^4(\vec{k}_{1,\perp}+ \vec{q}_{\perp})^2}\nn\\
			&\times\int \frac{d^4k_2}{(2\pi)^4 k_{2,\perp}^2}\frac{|2k_2^z-2k_1^z|^{-\eta}|2k_2^z|^{-\eta}}{(\vec{k}_{2,\perp}-\vec{k}_{1,\perp})^2}\frac{iQ^-}{(Q-k_1-q)^2+i\epsilon}\frac{iQ^-}{(Q-k_1+k_2-q)^2+i\epsilon}\nn\\
			&\times p^+\left( \frac{i}{(p+k_1+q)^2+i\epsilon}-2\pi \delta((p+k_1+q)^2)n|p^0+k_1^0+q^0|\right) \nn\\
			&\times p^+\left( \frac{i}{(p+k_2+q)^2+i\epsilon}-2\pi \delta((p+k_2+q)^2)n|p^0+k_2^0+q^0|\right) 
    \end{align}
		\normalsize
		 \end{minipage}%
\end{figure}
We can first apply the power counting for our Glauber mode and simplify this expression 
\small
\begin{align}
&M^{(3)}_c= (-2g^2)^3T^{A_1}T^{A_2}T^{A_3} \otimes T^{A_2}T^{A_1}T^{A_3} S_{ns}\nu^{3\eta} \int \frac{d^4k_1|2k_1^z|^{-\eta}}{(2\pi)^4(\vec{k}_{1,\perp}+ \vec{q}_{\perp})^2}\int \frac{d^4k_2}{(2\pi)^4 k_{2,\perp}^2}\frac{|2k_2^z-2k_1^z|^{-\eta}|2k_2^z|^{-\eta}}{(\vec{k}_{2,\perp}-\vec{k}_{1,\perp})^2}\nn\\
			&\times\frac{i}{Q^+-k_1^+-q^+-\frac{(\vec{Q}_{\perp}+\vec{k}_{1,\perp}-\vec{q}_{\perp})^2}{Q^-}+\frac{i\epsilon}{Q^-}}\frac{i}{Q^+-k_1^++k_2^+-q^+-\frac{(\vec{Q}_{\perp}-\vec{k}_{1,\perp}+\vec{k}_{2,\perp}-\vec{q}_{\perp})^2}{Q^-}+\frac{i\epsilon}{Q^-}}\nn\\
			&\times\Bigg\{\frac{i}{p^-+k_1^-++q^--\frac{(\vec{p}_{\perp}+\vec{k}_{1,\perp}+\vec{q}_{\perp})^2}{p^+}+\frac{i\epsilon}{p^+}}\nn\\
			&-2\pi \text{sign}(p^+)\delta\left(p^-+k_1^-++q^--\frac{(\vec{p}_{\perp}+\vec{k}_{1,\perp}+\vec{q}_{\perp})^2}{p^+}\right)n\Big|\frac{p^+}{2}+\frac{p^-+k_1^-+q^-}{2}\Big|\Bigg\}\nn\\
			&\times \Bigg\{ \frac{i}{p^-+k_2^-++q^--\frac{(\vec{p}_{\perp}+\vec{k}_{2,\perp}+\vec{q}_{\perp})^2}{p^+}+\frac{i\epsilon}{p^+}}\nn\\
			&-2\pi \text{sign}(p^+)\delta\left(p^-+k_2^-++q^--\frac{(\vec{p}_{\perp}+\vec{k}_{2,\perp}+\vec{q}_{\perp})^2}{p^+}\right)n\Big|\frac{p^+}{2}+\frac{p^-+k_2^-+q^-}{2}\Big|\Bigg\} 
\end{align}
\normalsize 
Due to the presence of the rapidity regulator, we have to do the integral over $k_1^0,k_2^0$. While $Q^- >0$, $p^+$ can be positive or negative. For $k_1^0$ we choose to close the contour in the lower half plane while for $k_2^0$ we close it in upper half plane. This gives us
\small 
\begin{align}
&M^{(3)}_c= (-2ig^2)^3T^{A_1}T^{A_2}T^{A_3} \otimes T^{A_2}T^{A_1}T^{A_3} S_{ns}\nu^{3\eta} \int \frac{d^2k_{1,\perp}dk_1^z|2k_1^z|^{-\eta}}{(2\pi)^3(\vec{k}_{1,\perp}+ \vec{q}_{\perp})^2}\int \frac{d^2k_{2,\perp}dk_2^z}{(2\pi)^3 k_{2,\perp}^2}\frac{|2k_2^z-2k_1^z|^{-\eta}|2k_2^z|^{-\eta}}{(\vec{k}_{2,\perp}-\vec{k}_{1,\perp})^2}\nn\\
			&\times\frac{i}{2k_1^z+\Delta_1+i\epsilon}\frac{i}{2k_1^z-2k_2^z+\Delta_2+i\epsilon}\nn\\
			&\left( \Theta(p^+)-\text{sign}(p^+)n\left(\Big|\frac{p^+}{2}+\frac{(\vec{p}_{\perp}+\vec{k}_{1,\perp}+\vec{q}_{\perp})^2}{2p^+}\Big|\right)\right)\left( -\Theta(-p^+)- \text{sign}(p^+)n\left(\Big|\frac{p^+}{2}+\frac{(\vec{p}_{\perp}+\vec{k}_{2,\perp}+\vec{q}_{\perp})^2}{2p^+}\Big|\right) \right)
\end{align}
\normalsize 
We can now shift $k_2^z \rightarrow -k_2^z+k_1^z$ to write 
\small
\begin{align}
&M^{(3)}_c= (-2ig^2)^3T^{A_1}T^{A_2}T^{A_3} \otimes T^{A_2}T^{A_1}T^{A_3} S_{ns}\nu^{3\eta} \int \frac{d^2k_{1,\perp}dk_1^z|2k_1^z|^{-\eta}}{(2\pi)^3(\vec{k}_{1,\perp}+ \vec{q}_{\perp})^2}\int \frac{d^2k_{2,\perp}dk_2^z}{(2\pi)^3 k_{2,\perp}^2}\frac{|2k_1^z|^{-\eta}|2k_2^z-2k_1^z|^{-\eta}}{(\vec{k}_{2,\perp}-\vec{k}_{1,\perp})^2}\nn\\
			&\times\frac{i}{2k_1^z+\Delta_1+i\epsilon}\frac{i}{2k_2^z+\Delta_2+i\epsilon}\nn\\
			&\left( \Theta(p^+)-\text{sign}(p^+)n\left(\Big|\frac{p^+}{2}+\frac{(\vec{p}_{\perp}+\vec{k}_{1,\perp}+\vec{q}_{\perp})^2}{2p^+}\Big|\right)\right)\left( -\Theta(-p^+)- \text{sign}(p^+)n\left(\Big|\frac{p^+}{2}+\frac{(\vec{p}_{\perp}+\vec{k}_{2,\perp}+\vec{q}_{\perp})^2}{2p^+}\Big|\right) \right)
\end{align}
\normalsize 
Then following \cite{Rothstein:2016bsq}, we use the result 

\begin{align}
\nu^{3\eta}\int \frac{dk_1^z}{2\pi}\frac{dk_2^z}{2\pi}\frac{|2k_1^z|^{-\eta}|2k_2^z-2k_1^z|^{-\eta}|2k_2^z|^{-\eta}|}{(2k_1^z+\Delta_1+i\epsilon)(2k_2^z+\Delta_2+i\epsilon)}= \frac{1}{4} \frac{1}{3!} 
\end{align}
which gives us our final result quoted in Eq.\ref{nine}.
\small 
\begin{align}
&M^{(3)}_c= 2\frac{(-ig^2)^3}{3!}T^{A_1}T^{A_2}T^{A_3} \otimes T^{A_2}T^{A_1}T^{A_3} S_{ns} \int \frac{d^2k_{1,\perp}}{(2\pi)^2(\vec{k}_{1,\perp}+ \vec{q}_{\perp})^2}\int \frac{d^2k_{2,\perp}}{(2\pi)^2 k_{2,\perp}^2}\frac{1}{(\vec{k}_{2,\perp}-\vec{k}_{1,\perp})^2}\nn\\
			&\left( \Theta(p^+)-\text{sign}(p^+)n\left(\Big|\frac{p^+}{2}+\frac{(\vec{p}_{\perp}+\vec{k}_{1,\perp}+\vec{q}_{\perp})^2}{2p^+}\Big|\right)\right)\left(-\Theta(-p^+)- \text{sign}(p^+)n\left(\Big|\frac{p^+}{2}+\frac{(\vec{p}_{\perp}+\vec{k}_{2,\perp}+\vec{q}_{\perp})^2}{2p^+}\Big|\right) \right)
\end{align}


\providecommand{\href}[2]{#2}\begingroup\raggedright\begin{thebibliography}{}

\end{thebibliography}\endgroup


\normalsize 

\begin{thebibliography}{99}


\bibitem{Vaidya:2020lih}
V.~Vaidya,
[arXiv:2010.00028 [hep-ph]].


\bibitem{Gyulassy:1993hr} 
  M.~Gyulassy and X.~n.~Wang,
  Nucl.\ Phys.\ B {\bf 420}, 583 (1994)
  [nucl-th/9306003].

\bibitem{Wang:1994fx} 
  X.~N.~Wang, M.~Gyulassy and M.~Plumer,
  Phys.\ Rev.\ D {\bf 51}, 3436 (1995)
  [hep-ph/9408344].

\bibitem{Baier:1994bd} 
  R.~Baier, Y.~L.~Dokshitzer, S.~Peigne and D.~Schiff,
  Phys.\ Lett.\ B {\bf 345}, 277 (1995)
  [hep-ph/9411409].

\bibitem{Baier:1996kr} 
  R.~Baier, Y.~L.~Dokshitzer, A.~H.~Mueller, S.~Peigne and D.~Schiff,
  Nucl.\ Phys.\ B {\bf 483}, 291 (1997)
  [hep-ph/9607355].

\bibitem{Baier:1996sk} 
  R.~Baier, Y.~L.~Dokshitzer, A.~H.~Mueller, S.~Peigne and D.~Schiff,
  Nucl.\ Phys.\ B {\bf 484}, 265 (1997)
  [hep-ph/9608322].

\bibitem{Zakharov:1996fv}  
  B.~G.~Zakharov,
  JETP Lett.\  {\bf 63}, 952 (1996)
  [hep-ph/9607440].

\bibitem{Zakharov:1997uu} 
  B.~G.~Zakharov,
  JETP Lett.\  {\bf 65}, 615 (1997)
  [hep-ph/9704255].

\bibitem{Gyulassy:1999zd} 
  M.~Gyulassy, P.~Levai and I.~Vitev,
  Nucl.\ Phys.\ B {\bf 571}, 197 (2000)
  [hep-ph/9907461].

\bibitem{Gyulassy:2000er} 
  M.~Gyulassy, P.~Levai and I.~Vitev,
  Nucl.\ Phys.\ B {\bf 594}, 371 (2001)
  [nucl-th/0006010].

\bibitem{Wiedemann:2000za} 
  U.~A.~Wiedemann,
  Nucl.\ Phys.\ B {\bf 588}, 303 (2000)
  [hep-ph/0005129].

\bibitem{Guo:2000nz} 
  X.~f.~Guo and X.~N.~Wang,
  Phys.\ Rev.\ Lett.\  {\bf 85}, 3591 (2000)
  [hep-ph/0005044].
 
\bibitem{Wang:2001ifa} 
  X.~N.~Wang and X.~f.~Guo,
  Nucl.\ Phys.\ A {\bf 696}, 788 (2001)
  [hep-ph/0102230].
  
\bibitem{Arnold:2002ja} 
  P.~B.~Arnold, G.~D.~Moore and L.~G.~Yaffe,
  JHEP {\bf 0206}, 030 (2002)
  [hep-ph/0204343].

\bibitem{Arnold:2002zm} 
  P.~B.~Arnold, G.~D.~Moore and L.~G.~Yaffe,
  JHEP {\bf 0301}, 030 (2003)
  [hep-ph/0209353].
  
\bibitem{Salgado:2003gb} 
  C.~A.~Salgado and U.~A.~Wiedemann,
  Phys.\ Rev.\ D {\bf 68}, 014008 (2003)
  [hep-ph/0302184].

\bibitem{Armesto:2003jh} 
  N.~Armesto, C.~A.~Salgado and U.~A.~Wiedemann,
  Phys.\ Rev.\ D {\bf 69}, 114003 (2004)
  [hep-ph/0312106].

\bibitem{Majumder:2006wi}
A.~Majumder, B.~M\"uller and S.~A.~Bass,
Phys.\ Rev.\ Lett.\  \textbf{99}, 042301 (2007)
[arXiv:hep-ph/0611135 [hep-ph]].

\bibitem{Majumder:2007zh}
A.~Majumder, B.~M\"uller and X.~Wang,
Phys.\ Rev.\ Lett.\  \textbf{99}, 192301 (2007)
[arXiv:hep-ph/0703082 [hep-ph]].

\bibitem{Neufeld:2008fi}
R.~Neufeld, B.~M\"uller and J.~Ruppert,
Phys.\ Rev.\ C \textbf{78}, 041901 (2008)
[arXiv:0802.2254 [hep-ph]].

\bibitem{Neufeld:2009ep}
R.~Neufeld and B.~M\"uller,
Phys.\ Rev.\ Lett.\  \textbf{103}, 042301 (2009)
[arXiv:0902.2950 [nucl-th]].

\bibitem{Arsene:2004fa} 
  I.~Arsene {\it et al.} [BRAHMS Collaboration],
  Nucl.\ Phys.\ A {\bf 757}, 1 (2005)
  [nucl-ex/0410020].
  
\bibitem{Back:2004je} 
  B.~B.~Back {\it et al.},
  Nucl.\ Phys.\ A {\bf 757}, 28 (2005)
  [nucl-ex/0410022].

\bibitem{Adams:2005dq} 
  J.~Adams {\it et al.} [STAR Collaboration],
  Nucl.\ Phys.\ A {\bf 757}, 102 (2005)
  [nucl-ex/0501009].

\bibitem{Adcox:2004mh} 
  K.~Adcox {\it et al.} [PHENIX Collaboration],
  Nucl.\ Phys.\ A {\bf 757}, 184 (2005)
  [nucl-ex/0410003].

\bibitem{Aad:2010bu} 
  G.~Aad {\it et al.} [ATLAS Collaboration],
  Phys.\ Rev.\ Lett.\  {\bf 105}, 252303 (2010)
  [arXiv:1011.6182 [hep-ex]].

\bibitem{Aamodt:2010jd} 
  K.~Aamodt {\it et al.} [ALICE Collaboration],
  Phys.\ Lett.\ B {\bf 696}, 30 (2011)
  [arXiv:1012.1004 [nucl-ex]].

\bibitem{Chatrchyan:2011sx} 
  S.~Chatrchyan {\it et al.} [CMS Collaboration],
  Phys.\ Rev.\ C {\bf 84}, 024906 (2011)
  [arXiv:1102.1957 [nucl-ex]].
  




%
%
  %
%
%
%
%
%
%
%
%

\bibitem{Ovanesyan:2011kn} 
  G.~Ovanesyan and I.~Vitev,
  Phys.\ Lett.\ B {\bf 706}, 371 (2012)
  [arXiv:1109.5619 [hep-ph]].
  
  
\bibitem{Chien:2015hda} 
  Y.~T.~Chien and I.~Vitev,
  JHEP {\bf 1605}, 023 (2016)
  [arXiv:1509.07257 [hep-ph]].

\bibitem{Ovanesyan:2011xy} 
  G.~Ovanesyan and I.~Vitev,
  JHEP {\bf 1106}, 080 (2011)
  [arXiv:1103.1074 [hep-ph]].

\bibitem{Chien:2015vja} 
  Y.~T.~Chien, A.~Emerman, Z.~B.~Kang, G.~Ovanesyan and I.~Vitev,
  Phys.\ Rev.\ D {\bf 93}, no. 7, 074030 (2016)
  [arXiv:1509.02936 [hep-ph]].

\bibitem{Kang:2014xsa} 
  Z.~B.~Kang, R.~Lashof-Regas, G.~Ovanesyan, P.~Saad and I.~Vitev,
  Phys.\ Rev.\ Lett.\  {\bf 114}, no. 9, 092002 (2015)
  [arXiv:1405.2612 [hep-ph]].

\bibitem{Rothstein:2016bsq}
I.~Z.~Rothstein and I.~W.~Stewart,
JHEP \textbf{08}, 025 (2016)
[arXiv:1601.04695 [hep-ph]].
  
%
%
%
%
%
  %
%
%
%
%
 %
%
  %
%
%
%
  %
%
%
%
      %
  %
%
  %
%
  %
  %
%
  %
  
\bibitem{Vaidya:2020cyi}
V.~Vaidya and X.~Yao,
[arXiv:2004.11403 [hep-ph]].
	
	%
	
	
\bibitem{Chiu:2012ir}
J.~Y.~Chiu, A.~Jain, D.~Neill and I.~Z.~Rothstein,
JHEP \textbf{05}, 084 (2012)
doi:10.1007/JHEP05(2012)084
[arXiv:1202.0814 [hep-ph]].
	
%
%
%
%
%
%


\bibitem{Bauer:2000yr}
C.~W.~Bauer, S.~Fleming, D.~Pirjol and I.~W.~Stewart,
Phys.\ Rev.\ D \textbf{63}, 114020 (2001)
[arXiv:hep-ph/0011336 [hep-ph]].


\bibitem{Bauer:2001ct}
C.~W.~Bauer and I.~W.~Stewart,
Phys.\ Lett.\ B \textbf{516}, 134-142 (2001)
[arXiv:hep-ph/0107001 [hep-ph]].
  
  
\bibitem{Bauer:2002nz}
C.~W.~Bauer, S.~Fleming, D.~Pirjol, I.~Z.~Rothstein and I.~W.~Stewart,
Phys.\ Rev.\ D \textbf{66}, 014017 (2002)
[arXiv:hep-ph/0202088 [hep-ph]].


%
%
%
%
%
%
%
%
%


\end{thebibliography}
\end{document}